\documentclass[reprint,  amsmath, amssymb, aps]{revtex4-2}

\usepackage{amssymb,amsmath,amsfonts}
\usepackage{graphicx,graphics}
\usepackage{appendix}
\usepackage{epsfig}
\usepackage{epstopdf}
\usepackage{amsfonts}
\usepackage{bm,bbm}
\usepackage{amssymb,amsmath,amsfonts}
\usepackage{mathrsfs}
\usepackage{calc}
\usepackage{epsfig}
\usepackage{float}
\usepackage{xcolor}
\usepackage{color}
\usepackage{epstopdf}
\usepackage{bm,amssymb,amsmath}
\usepackage{graphicx,color}

\usepackage{ulem}

\begin{document}

\title{Resetting as a swift equilibration protocol in an anharmonic potential}
\author{R\'emi Goerlich}
\affiliation{Raymond \& Beverly Sackler School of Chemistry, Tel Aviv University, Tel Aviv 6997801, Israel}
\author{Tommer D. Keidar}
\affiliation{Raymond \& Beverly Sackler School of Chemistry, Tel Aviv University, Tel Aviv 6997801, Israel}
\author{Yael Roichman}
\email{roichman@tauex.tau.ac.il}
\affiliation{Raymond \& Beverly Sackler School of Physics and Astronomy, Tel Aviv University, Tel Aviv 6997801, Israel}
\affiliation{Raymond \& Beverly Sackler School of Chemistry, Tel Aviv University, Tel Aviv 6997801, Israel}

\date{\today}
\begin{abstract}
We present and characterize a method to accelerate the relaxation of a Brownian object between two distinct equilibrium states. 
Instead of relying on a deterministic time-dependent control parameter, we use stochastic resetting to guide and accelerate the transient evolution.
The protocol is investigated theoretically, and its thermodynamic cost is evaluated with the tools of stochastic thermodynamics.
Remarkably, we show that stochastic resetting significantly accelerates the relaxation to the final state. This stochastic protocol exhibits energetic and temporal characteristics that align with the scales observed in previously investigated deterministic protocols. Moreover, it expands the spectrum of stationary states that can be manipulated, incorporating new potential profiles achievable through experimentally viable protocols.
\end{abstract}

\maketitle

Swift driving protocols in non-equilibrium statistical mechanics aim to manage a system's transition between thermal equilibrium states efficiently, seeking quicker routes to equilibrium than natural relaxation \cite{patra_shortcuts_2017, guery-odelin_shortcuts_2019, guery-odelin_driving_2023}. For instance, various techniques were used to showcase a reduction in the relaxation time ($\tau_r$) of a Brownian particle coupled to a thermal bath after a sudden change in a control parameter, e.g., potential stiffness or temperature. Such swift state-to-state transformation (SST) protocols usually rely on the full knowledge of the time-dependant probability density and its relation to external control parameters.
This allows reverse engineering a well-chosen temporal change of the control parameter ending at the new equilibrium state \cite{guery-odelin_driving_2023}.
In the case of harmonic confinement, explicit SST protocols were derived and experimentally implemented to control the relaxation of Brownian particles \cite{martinez_engineered_2016, Chupeau2018, Raynal2023} with promising applications such as directly increasing the efficiency of microscopic heat engines \cite{blickle2012realization, martinez2016brownian, argun2017experimental, chiang2017electrical, martinez2017colloidal}.

Such acceleration beyond thermal relaxation comes at a cost.
The energetic expenses of these control protocols are effectively analyzed through stochastic thermodynamics at the level of individual stochastic trajectories \cite{Sekimoto1998, seifert2012}.
Therefore, general principles governing such protocols' trade-offs between time and energy are calculated, allowing for their optimization. \cite{Schmiedl2007, zhang_work_2020, Rosales2020, Pires2023,loos2024}.

The current focus of mesoscopic thermodynamics on inherently non-equilibrium systems like active matter is challenging the existing SST methods \cite{krishnamurthy2016micrometre, martin2018extracting, baldovin2023control}.
This led recently to the derivation of SST between non-equilibrium states \cite{baldassarri_engineered_2020, Prados2021} and for arbitrary potentials \cite{Plata2021}, which are more complex and delicate to implement experimentally. This is the motivation of our work, seeking experimentally applicable SST methods in non-equilibrium and non-harmonic frameworks.

Stochastic resetting (SR) is a driving mechanism in which a process is arrested randomly only to be re-initiated repeatedly, usually from the origin. The classic example of SR is resetting the motion of a diffusing Brownian particle \cite{evans_diffusion_2011, evans_stochastic_2020}, which has been realized experimentally employing optically trapped Brownian colloidal particles \cite{tal2020, besga2020, goerlich2023}, and thoroughly investigated theoretically \cite{evans_diffusion_2011, evans2011diffusion, evans2013optimal, pal2017first, chechkin2018random, Gupta2022}, demonstrating fluctuation-dissipation relations \cite{sokolov2023linear}, enhanced sampling \cite{blumer2022stochastic} or SR-induced Mpemba effect \cite{busiello2021inducing}.
The rising interest in SR is prompted, among other features of SR, by the emergence of a stationary state, making SR akin to confinement yet bearing unique characteristics. 
Since each resetting event in SR breaks detailed balance, the resulting steady state is, of course, out of thermal equilibrium, with non-vanishing probability currents and constant dissipation of heat to the surrounding fluid \cite{fuchs_stochastic_2016, mori_entropy_2023, Gupta2022, goerlich2023}.

Here, we propose an innovative method to accelerate the relaxation between two states based on SR rather than on a controlled variation of the underlying potential. In our method, the potential is switched off during the transition, and the system undergoes stochastic resetting until its position distribution reaches that of the target state. Subsequently, the target potential is turned on, and the stochastic resetting is terminated.
The acceleration is rooted in the detailed-balance breaking jumps intrinsic to SR, explicitly using the non-equilibrium nature of the process to expedite the relaxation.

Specifically, we demonstrate that stochastic resetting accelerates the transition of a Brownian particle between two equilibrium states with a v-shaped potential. Moreover, it naturally connects two non-equilibrium steady-states where standard SSTs are difficult to derive.
This method shares similarities with a previous proposal \cite{Roldan2017} where an energy-dependant SR rate, contingent upon knowledge of particle positions in time, expedites transitions between two Gaussian equilibrium states. In sharp contrast, the method developed here does not require such knowledge. Finally, we characterize the entropic cost of accelerated relaxation protocols via stochastic thermodynamics and discuss an extension to finite-time resetting, offering experimentally feasible acceleration.

\begin{figure}[htb!]
    \centering
    \includegraphics[width=1\linewidth]{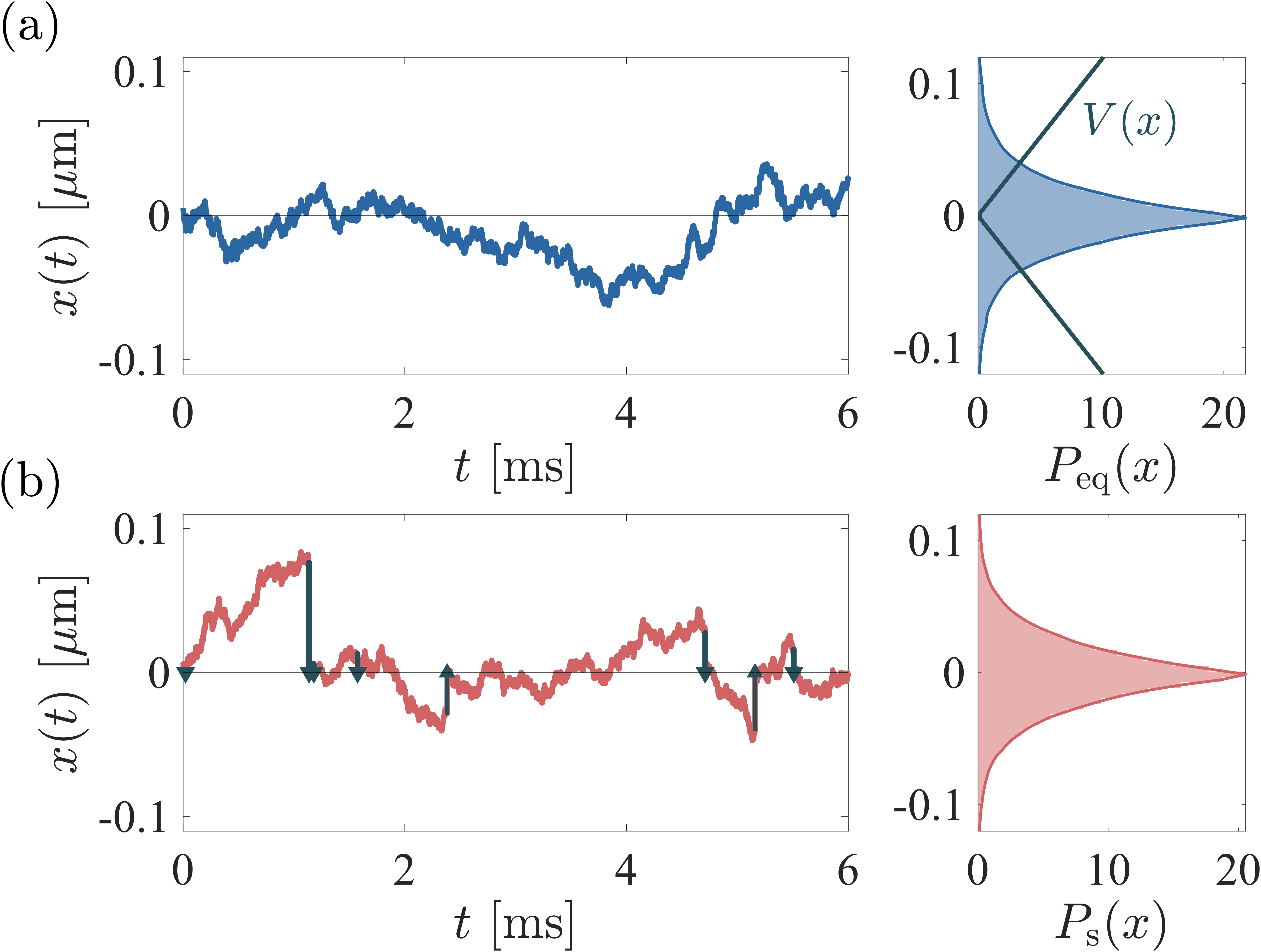}
    \caption{(a) Time-series of position $x(t)$ obeying the Langevin process Eq.~(\ref{Eq:Langevin}) at thermal equilibrium in a v-shaped potential with slope $b = 0.18 $ pN; the Maxwell-Boltzmann equilibrium distribution is plotted on the right.
    (b) Time-series of positions $x(t)$ Brownian particle undergoing stochastic resetting with $\lambda = b^2/\gamma k_BT = 1$ kHz (red line). Resetting events are shown with vertical arrows, and the Laplace PDF is plotted on the right.
    }
    \label{fig:Traj}
\end{figure}

\textit{Tailoring identical position distribution for equilibrium and driven systems - } We start by considering a Brownian particle diffusing in a confining linear v-shaped potential $V(x) = b |x|$ with $b > 0$, obeying the Langevin equation,
\begin{equation}
    \dot{x}(t) = \frac{-b}{\gamma} \textrm{sgn} (x(t)) + \sqrt{2 D} \xi(t),
    \label{Eq:Langevin}
\end{equation}
where $\gamma$ is the viscous drag coefficient, $\textrm{sgn}(x(t))$ is the sign function, $D = k_B T / \gamma$ is the diffusion coefficient with $k_B$ is Boltzmann's constant and $T$ the temperature.
$\xi(t)$ is a Gaussian random variable with $\langle \xi(t) \rangle = 0$ and $\langle \xi(t) \xi(s) \rangle = \delta(t-s)$. The position probability density function (PDF) of the particle, according to Boltzmann statistics, is given by,
\begin{equation}
    P_{\rm eq}(x) = \frac{b}{2k_BT} e^{-\frac{b}{k_BT} |x|}.
    \label{Eq:DistEq}
\end{equation}
An exemplary Langevin simulation of such a particle is depicted in Fig.~\ref{fig:Traj} (a), together with its stationary position distribution. Throughout the paper, we consider a micron-sized colloidal particle at room temperature $T = 300$ K with $\gamma = 8 \times 10^{-9} ~ \rm{kg}~\rm{s}^{-1}$.
All simulations are initialized with random positions in the steady-state, eliminating an initial thermalization time.
The PDFs (here and below) are measured on a single trajectory of $10^7$ time-steps of duration $dt = 10^{-6}$ s.

In the absence of an underlying potential, SR of a Brownian particle to the origin $x=0$ with resetting times drawn from an exponential distribution, $P(t_{\rm rst}) = \lambda e^{-\lambda t_{\rm rst}}$, results in the following stationary position PDF, 
\begin{equation}
    P_s(x) = \frac{\alpha}{2} e^{-\alpha |x|}
    \label{Eq:DistSr}
\end{equation}
where $\alpha = \sqrt{\lambda / D}$. A typical trajectory of such a process alongside its PDF is shown in Fig.~\ref{fig:Traj} (b).

Central to our proposed acceleration protocol of an SST is the fact that this probability distribution (Eq.~\ref{Eq:DistSr}) can be tuned to be identical to the equilibrium distribution of Eq.~\ref{Eq:DistEq} by taking $b=\alpha k_BT$. Namely, by fixing the resetting rate to be $\lambda_b=b^2/\gamma k_BT$.

\begin{figure}[htb!]
    \centering
    \includegraphics[width=1\linewidth]{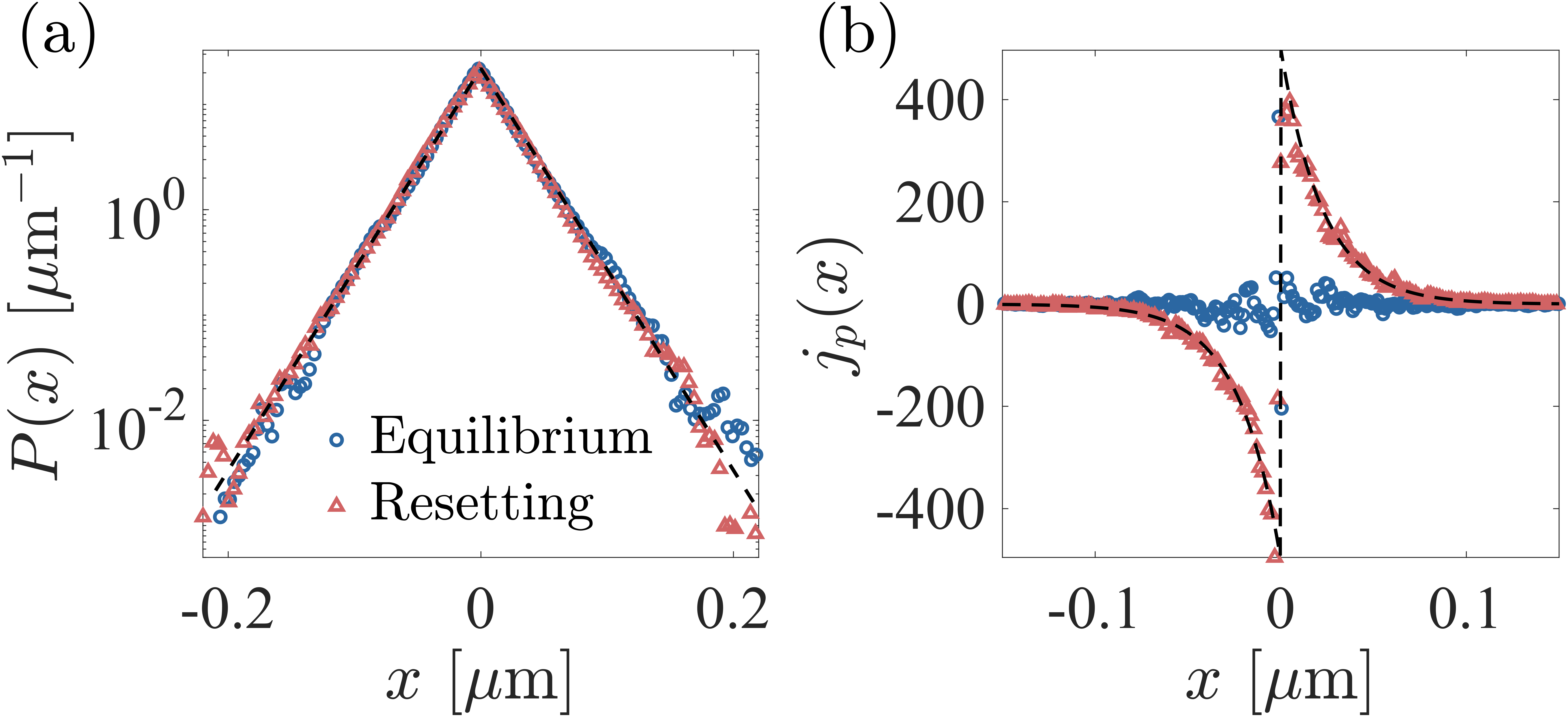}
    \caption{(a) Probability distribution of an equilibrium process (blue circles) and a resetting process (red triangles), along with the analytical result $P_{\rm eq}(x) = P_s(x)$ (black dashed line) for a single trajectory.
    (b) probability density current through space for the equilibrium state (blue circles) and for SR (red triangles) together with the analytical result for SR (black dashed line).}
    \label{fig:Dist}
\end{figure}

In Fig.~\ref{fig:Dist}(a), we show quantitatively that by installing the appropriate resetting rate $\lambda_b$ we obtain the expected identical Laplace PDFs for the two cases: equilibrium with a V-shapedpotential (Eq.~\ref{Eq:DistEq}) and Poissonian stochastic resetting (Eq.~\ref{Eq:DistSr}).
Importantly, having identical PDFs does not imply similar dynamics. Under SR, a particle experiences significant jumps at each resetting event, leading to a non-zero stationary current $j_p(x) =  \frac{1}{\gamma}\frac{d V(x)}{d x}P_s(x) - D \frac{d P_s(x)}{d x} = \frac{D \alpha^2}{2} \rm{sgn}(x) e^{-\alpha |x|}$, as illustrated in Fig.~\ref{fig:Dist}(b). This stands in stark contrast to the behavior of a particle at equilibrium. The distinct dynamics under SR enable accelerating a transition between stationary states.

\textit{Accelerating state-to-state transition - }
The simplest unassisted transition between two equilibrium states involves an abrupt change of the potential slope from $b(t_i)$ to $b(t_f)$ followed by spontaneous relaxation.
Similarly, a freely diffusing particle in steady state under resetting can be subjected to a sudden change of the resetting rate from $\lambda_b(t_i)=b(t_i)^2/\gamma k_BT$ to $\lambda_b(t_f)=b(t_f)^2/\gamma k_BT$ followed by a relaxation to the new steady state.

\begin{figure}[htb!]
    \centering
    \includegraphics[width=1\linewidth]{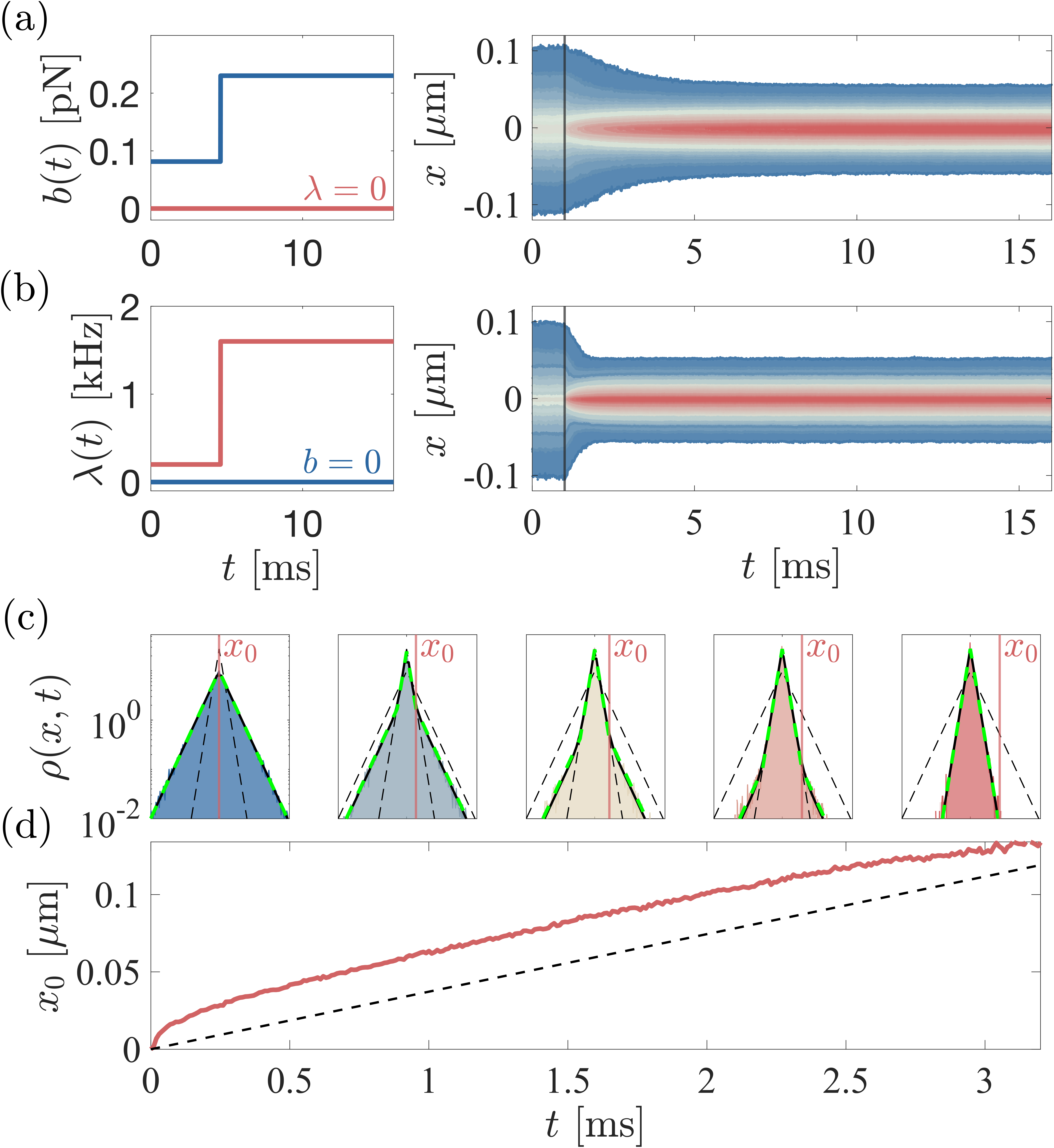}
    \caption{
    (a) Step-like change of the potential slope $b(t)$ in the absence of resetting ($\lambda = 0$) and corresponding contour plot of the  time-dependent probability density.
    (b) Step-like change in stochastic resetting rate $\lambda(t)$ in the absence of potential ($b = 0$) and corresponding time-dependent probability density.
    (c) five distinct PDF (from blue to red) during the transient evolution of the same ensemble under SR-assisted SST, the analytical solution $\rho_{\rm exact}(x,t)$, computed from the Laplace-space solution (black lines) and the empirical profile $f(x,t)$ (green dashed lines) using the values $x_0(t)$ extracted from a fitting procedure.
    (d) time-dependent boundary $x_0(t)$ (red line) up to the limit defined by numerical precision together with the theoretical evolution (black dashed line) from $\delta$ initial conditions derived in \cite{Majumdar2015} using a slope $2(\sqrt{D \lambda(t_f)} - \sqrt{D \lambda(t_i)})$.
    }
    \label{fig:ContourProtocol}
\end{figure}

In Fig.~\ref{fig:ContourProtocol}(a, b)  simulation results of both protocols are compared by depicting the evolution of the PDF, for a  potential quench (a), and for a SR-assisted transition (b). The position distribution is obtained from an ensemble average of $10^5$ independent trajectories of 2800 time-steps with $dt = 10 \mu \rm{s}$. While the initial and final states of both protocols have the same PDF, the rate of its evolution is different.

\begin{figure}[htb!]
    \centering
    \includegraphics[width=1\linewidth]{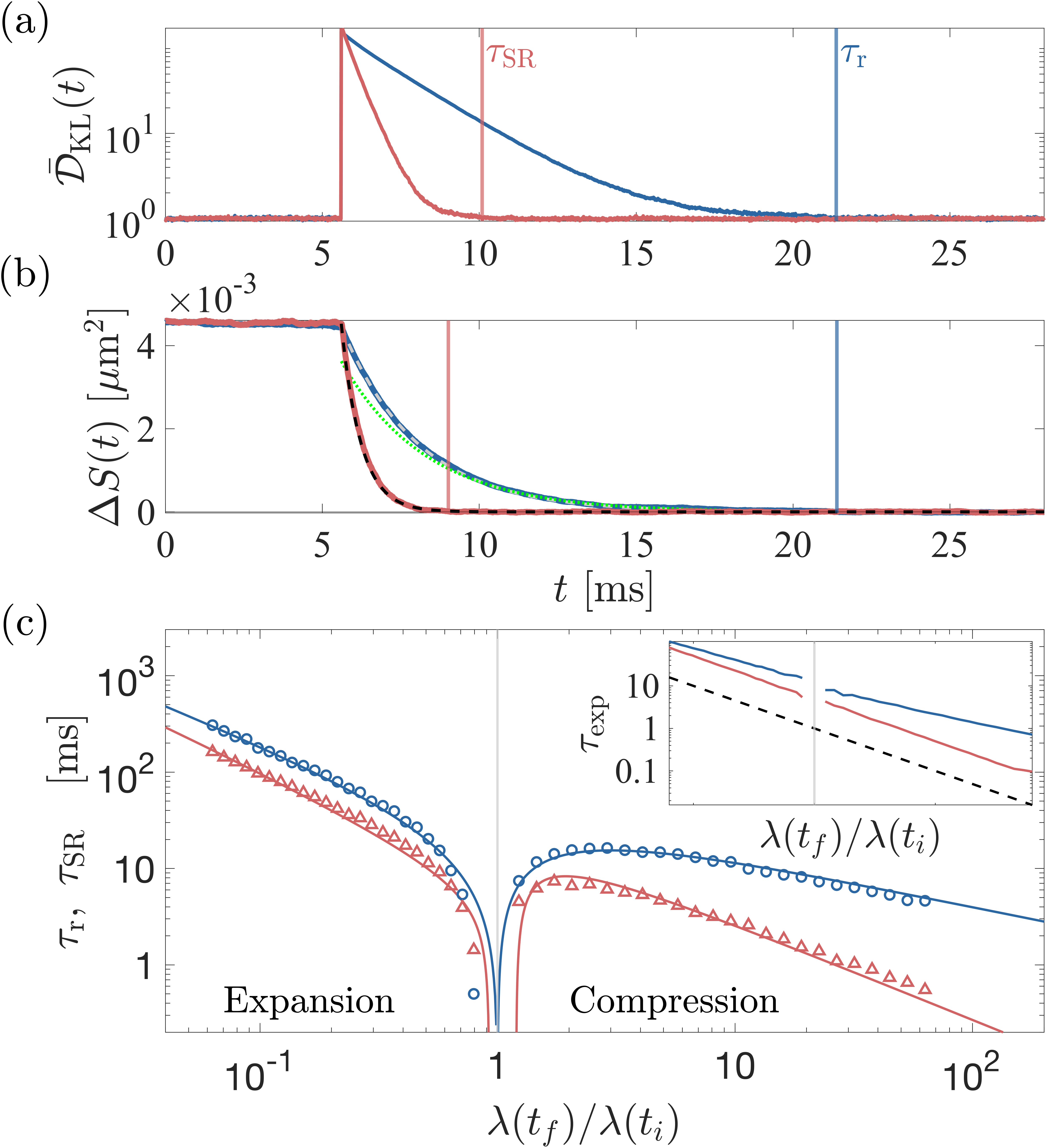}
    \caption{
    (a) Relative Kullback-Leibler divergence as a measure in natural bits of the distance to stationary state, normalized by the stationary value. The response to a potential quench is shown with a blue line and the SR-assisted transition as a red line. Relaxation times are underlined with blue and red vertical lines.
    (b) Time-dependent variance $\Delta S(t) = \langle x^2(t) \rangle - \frac{2D}{\lambda(t_f)}$ decay in both cases. Together with Eq.~(\ref{eq:var}) (black dashed line). For the potential quench, we show both a simple exponential fit (grey dashed line) and the long-time approximation of the exact result derived in Appendix \ref{App:Variance} (green dashed line).
    (c) Measured relaxation times $\tau_r$ (blue circles) and $\tau_{\rm{SR}}$ (red triangles) as a function of the ratio of compression/expansion $\lambda(t_f)/\lambda(t_i)$. Here, the initial state is defined by $\lambda(t_i) = 200$ Hz.  
    The superimposed lines correspond to theoretical characteristic times as detailed in the main text.
    The characteristic exponential relaxation times $\tau_{\rm exp}$ measured by fitting the variance relaxations with a simple exponential, are shown in the inset, spanning over the same range of $\lambda(t_f)/\lambda(t_i)$. For the quench in resetting rate (red line), we observe $ \tau_{\rm exp} \propto \lambda(t_i)/\lambda(t_f)$ (black dashed line).
    }
    \label{fig:KlAndTimes}
\end{figure}

The detailed evolution of the PDF after a quench of resetting rate $\lambda$ is shown on Fig.~\ref{fig:KlAndTimes}(c) with histograms spanning across the relaxation from blue to red.
Clearly, the PDF splits into two regions, sharply separated by a front that propagates from the center to the tails.
This motivates us to generalize the solution of  \cite{Majumdar2015}, where the time-dependent PDF is derived for a $\delta(x)$-distributed initial condition by integrating it over the initial steady-state $P_i(x) = \alpha(t_i) e^{-\alpha(t_i)|x|}/2$.
This integral can be solved in Laplace-domain (see Appendix \ref{App:PDF}), yielding the solution,
\begin{align*}
	\tilde{P}(x, s) &= \frac{\Delta \lambda \sqrt{s + \lambda(t_f)}}{2 s \sqrt{D}(\Delta \lambda - s)} e^{-\sqrt{\frac{s + \lambda(t_f)}{D}} |x|}\\
	& - \frac{\sqrt{\lambda(t_i)}}{2\sqrt{D}(\Delta \lambda - s)} e^{-\alpha(t_i) |x|},
\end{align*}
where $s$ is the Laplace-space variable and $\Delta \lambda \equiv \lambda(t_f) - \lambda(t_i)$.
The exact time-dependent PDF, $\rho_{\rm exact}(x,t)$ obtained by numerical inversion is shown in Fig. \ref{fig:ContourProtocol} (c), agrees well with the numerical PDF measured on independent stochastic processes.

A time-dependant front $x_0(t)$ splits an inner core region where $\rho(|x|<x_0,t) = P_f(x) = \alpha(t_f) e^{-\alpha(t_f) |x|}/2$, from an the outer region where  $\rho(|x|>x_0,t) = P_i(x) = \alpha(t_i) e^{-\alpha(t_i)|x|}/2$.
This is similar to the result obtained in \cite{Majumdar2015}, where the front was interpreted as a dynamical phase transition separating stochastic trajectories that have undergone resetting to those that have not yet.

This very simple form of the distribution motivates us to propose an empirical expression for the PDF
\begin{equation}
    f(x,t) = \bar{w}(x,t) P_i(x) + w(x,t) e^{-[\alpha(t_f) - \alpha(t_i)] x_0(t)} P_f(x),
    \label{eq:TimeDepPDFMain}
\end{equation}
with $w$ a window-function obeying $w(t) = 1$ for $|x| < x_0(t)$ and $0$ for $|x| > x_0(t)$, while $\bar{w} = 1 - w$.
This result is shown on Fig.~\ref{fig:KlAndTimes}(c) and agrees well with the numerically measured $\rho(x,t)$ as well as the exact solution $\rho_{\rm exact}(x,t)$.
The time-dependent value of the front $x_0(t)$ (see Fig.~\ref{fig:ContourProtocol} (d)) is obtained by fitting the measured PDF to $f(x,t)$ during the transient (as detailed in Appendix \ref{App:PDF} ).
After a transient time due to the initial condition ($P_i(x)\neq\delta(x)$),  $x_0(t)$ evolves linearly in time (long-time errors are detailed in Appendix \ref{App:PDF}).
Importantly, the algebraic growth of the dynamical phase transition $x_0(t)$ prevents the definition of an unambiguous relaxation time; this motivates us to define relaxation as the time needed for $\rho(x,t)$ to be indistinguishable of $P_s(x)$ within a finite precision.
We therefore quantify the acceleration induced by SR (Fig.~\ref{fig:ContourProtocol} (a, b)), via the Kullback-Leibler (KL) divergence between the measured $\rho(x,t)$, and the stationary solution $P_s(x) = P_{\rm eq}(x)$ with the control-parameter $b(t)$ or $\lambda_b(t)$
\begin{equation}
	\mathcal{D}_{\rm KL}(\rho(x,t) \parallel P_s(x)) = \int \rho(x,t) \ln \left[ \frac{\rho(x,t)}{P_s(x)} \right]dx.
 \label{eq:LaplaceDist}
\end{equation}
When the control parameter is abruptly modified, $P_s(x)$ immediately adapts, while the system responds gradually, characterized by $\rho(x,t)$. The delay in the system's response is well captured by $ \bar{\mathcal{D}}_{\rm KL} = \mathcal{D}_{\rm KL} (t) / \mathcal{D}_{\rm KL, s}$, a normalized measure with respect to the stationary-state value $\mathcal{D}_{\rm KL, s}$ (Fig.~\ref{fig:KlAndTimes} (a)). The latter is only related to the numerical precision of the histogram. 

To measure numerically the relaxation times (Fig.~\ref{fig:KlAndTimes} (a), blue and red vertical lines), $\tau_\text{SR}$ and $\tau_\text{r}$, we probe the first value of the time-dependent $\bar{\mathcal{D}}_{\rm KL}$ below a the average final steady-state value.
As seen on Fig.~\ref{fig:KlAndTimes} (a) with a reasonable numerical precision, this corresponds effectively to the full relaxation of the distribution.

Remarkably, we observe a significantly faster relaxation to the new steady state for the SR-assisted SST, with $\tau_{\rm{SR}} = 4.49 $ ms, compared to its equilibrium counterpart, $\tau_{\rm{r}} = 15.8 $ ms (Fig.~\ref{fig:KlAndTimes}, (a)). In other words, the non-equilibrium steady state (NESS) evolution induced by stochastic resetting is more than three times faster than the relaxation from equilibrium to equilibrium under the selected parameters.

The evolution of the KL-divergence fully characterizes the decay of the PDF, namely it includes all moments.
However, useful information can be obtained by looking at the evolution of the second moment $S(t) = \langle x^2(t) \rangle$ (here, $\langle x \rangle =0$).
From the expression of the PDF in Laplace-space Eq.~(\ref{eq:LaplaceDist}) we can obtain a simple analytical form for the time-dependant variance under a change of resetting rate (see appendix \ref{App:Variance}),
\begin{equation}
    S(t) = 2 D \left( \frac{1}{\lambda(t_i)} - \frac{1}{\lambda(t_f)}\right) e^{-\lambda(t_f) t} + \frac{2D}{\lambda(t_f)}.
    \label{eq:var}
\end{equation}
This exponential decay (black dashed line in Fig.~\ref{fig:KlAndTimes}(b)), fully characterized by the final resetting rate, shows perfect agreement with the numerically measured variance relaxation (red line).

Similarly, for a transition between equilibrium states in V-shaped potential, $S(t)$ can be fitted with an exponential decay (Fig. \ref{fig:KlAndTimes} (b), blue line and grey dashed line).
The long-time limit of the variance can also be obtained from the exact propagator (Appendix \ref{App:VarianceVp}) and is shown on Fig. \ref{fig:KlAndTimes} (b) as a green dotted line.
We note that for many practical applications, the relaxation of the variance can be a sufficient criterion. However, here, we treat the most general case and base our quantitative description of the relaxation on the KL-divergence.

The SR-based acceleration method remains valid across the entire spectrum of potential changes, which we demonstrate by measuring the KL-divergence relaxation time while varying the final value of $\lambda(t_f)$ (respectively $b(t_f)$) with fixed $\lambda(t_i) = 200$ Hz ($b(t_i) = 0.08$ pN).
The measured relaxation times are shown in Fig.~\ref{fig:KlAndTimes} (c).
The NESS-to-NESS transition (red) is always faster than the equivalent potential quench (blue) for both compression and expansion.

As detailed in appendix \ref{App:detailsMoments}, Eq.~(\ref{eq:var}) allows to derive the time needed for the variance to converge arbitrarily close (up to some tolerance parameter) to its final steady-state value.
This time takes a simple form $\propto \alpha + \ln (1 - \lambda(t_f) / \lambda(t_i))$ where $\alpha$ depends both on the tolerance parameter and the precision of the simulation.
We show in Appendix \ref{App:detailsMoments} that this form captures perfectly the behaviour of the variance, but remarkably, it also provides a good approximation for the relaxation of the KL-divergence.
This is shown on Fig.~\ref{fig:KlAndTimes} (c) where we use the aforementioned expression to fit the measured relaxation time (red solid line).

The nature of the relaxation is different for compression and expansion (Appendix \ref{App:detailsMoments}).
When the potential is expanded (left half of Fig.~\ref{fig:KlAndTimes} (c)), the relaxation is dominated by diffusive forces. This is manifested in the hybrid shape of the PDF, which is a combination of Laplace and Gaussian shapes (Fig. \ref{fig:HistCompExp}).
We attribute the Laplace shape to conservative forces and the Gaussian shape to diffusion. The shape of this PDF is fully captured by the exact result Eq.~(\ref{eq:PdfVp}).

Being dominated by diffusion, we expect the transient time for expansion to increase monotonically with the difference between the two states. This difference is captured by, $\propto \Delta \mathcal{T} = |\mathcal{T}_f - \mathcal{T}_i|$, where the  typical timescale arises naturally as $\mathcal{T} =\frac{k_B T \gamma}{b^2} = \lambda_b^{-1}$.
The relaxation times for an expansion of a V-shaped potential shown in Fig.~\ref{fig:KlAndTimes} (c) are fitted with a simple proportionality factor to $ \Delta \mathcal{T}$.

In contrast, for compression (right half of Fig.~\ref{fig:KlAndTimes} (c)), the evolution is dominated by the applied conservative force, yielding a bi-Laplace shaped PDF (Fig.~\ref{fig:ContourProtocol} and Fig.~\ref{fig:HistCompExp}).
There is, therefore, a competition between the acceleration due to the increasing conservative force applied $\propto b(t_f)$ and the growing difference between initial and final states.
These effects are combined in a phenomenological fit $\tau_r \propto \frac{b(t_i)}{b(t_f)}\Delta \mathcal{T}$ with a single proportionality factor (Fig.~\ref{fig:KlAndTimes} (c)).
Naturally, in the limit of $\lambda(t_f)\rightarrow\lambda(t_i)$ the transition becomes infinitely fast for any protocol (seen as a cusp in Fig.~\ref{fig:KlAndTimes}(b)).
We stress again that the relaxation is defined via an arbitrary threshold.
Therefore, the value of the proportionality factors in the fits should not be considered as physically meaningful. Nonetheless, it gives an estimate of the time needed to approach a state that is practically indistinguishable from the fully relaxed state. 
Overall, these results clearly demonstrate the ability of SR to accelerate the relaxation of a thermal system via SR NESS-to-NESS transformation.

\begin{figure}[htb!]
    \centering
    \includegraphics[width=1\linewidth]{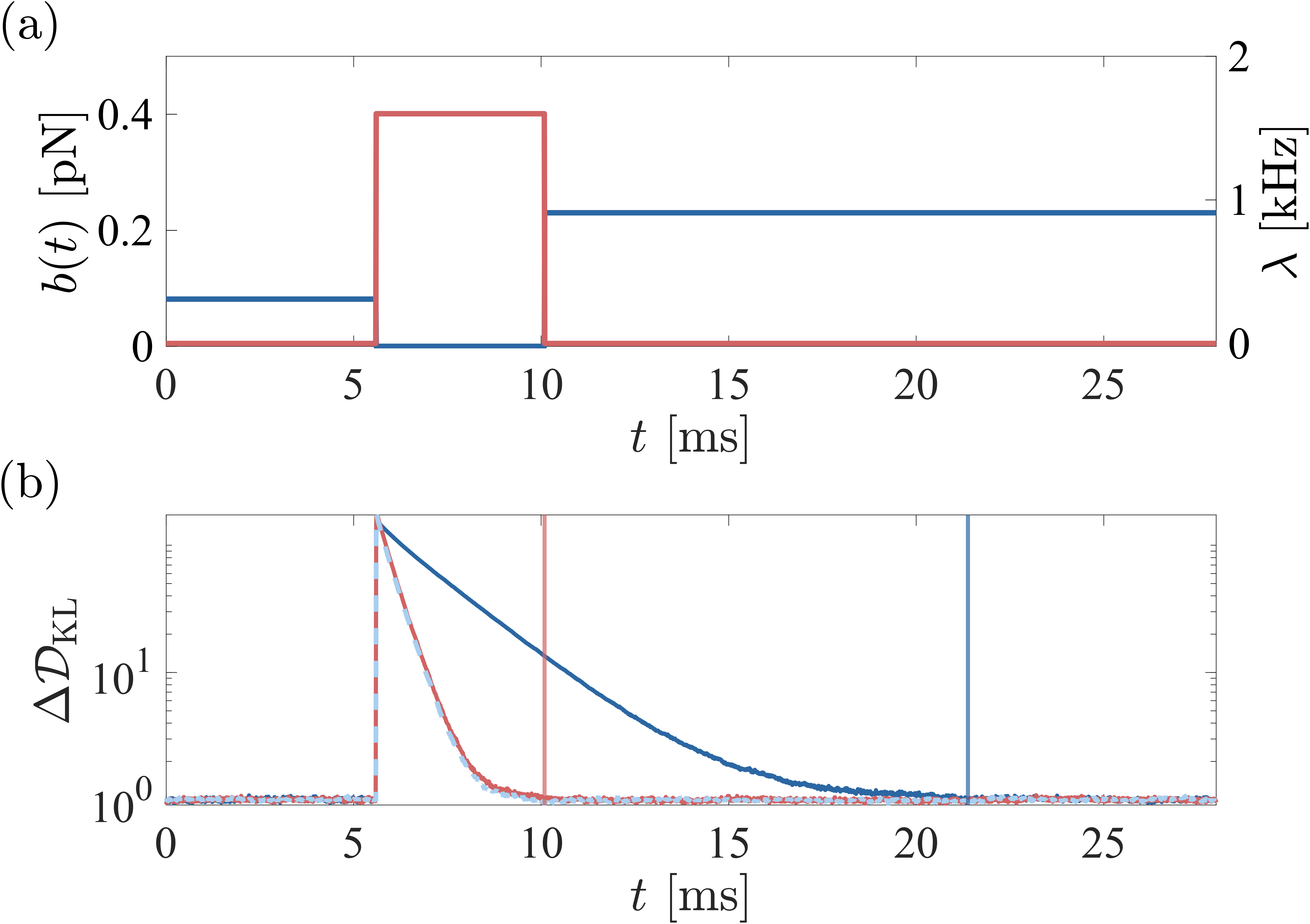}
    \caption{
    (a) schematic representation of the proposed protocol: initially the system is at thermal equilibrium in  $V(x) = b(t_i) |x|$, then (first black vertical line) the potential is replaced by SR with $\lambda = b(t_f)^2/(k_B T \gamma)$ for a time $\tau_{\rm{SR}}$ and finally (second black vertical line) resetting is turned off and the system is at equilibrium in $V(x) = b(t_f) |x|$.
    (b) $\Delta \mathcal{D}_{\rm KL}$ 
    for standard potential quench (deep blue line), SR-induced NESS-to-NESS (red line) and following our protocol (light blue dashed line).
    }
    \label{fig:Protocol}
\end{figure}

\textit{Accelerating relaxation to equilibrium and thermodynamic cost - }
The ability of SR to induce fast transitions between two NESS can be adapted to expedite transitions between two equilibrium states as well. SR is then used only during the transient between the two equilibrium states. The proposed protocol is comprised of the following steps (Fig.~\ref{fig:Protocol} (a)): \textit{(i)} a thermal equilibrium state in prepared in $V(x) = b(t_i) |x|$; \textit{(ii)} the potential is switched off and SR is applied with a rate $\lambda = b(t_f)^2/(k_B T \gamma)$ corresponding to the target state, \textit{(iii)} immediately after the system has relaxed (within the finite numerical precision) with time $\tau_{\rm{SR}} < \tau_r$, resetting is stopped and the potential is switched on in its final state $V(x) = b(t_f) |x|$.
Following this protocol fastens the particle's transition to the final equilibrium state compared to a potential quench protocol (Fig.~\ref{fig:Protocol} (b)) and relies on the  \textit{a priori} knowledge of $\tau_{\rm{SR}}$.

However, this approach relies on the transition from equilibrium to NESS (beginning of stage (\textit{ii})) and back (beginning of stage (\textit{iii})) to be instantaneous or with negligible time expenditure.
In other words, the PDFs dynamics following this protocol should be identical for transitions between NESS and equilibrium states, which is verified in Fig.~\ref{fig:Protocol}(b), where the dynamics of $\mathcal{D}_{\rm KL}$ of both SR-based transitions overlap and decay faster than that of the potential quench transition between equilibrium states.

Operating through markedly different mechanisms, both protocols entail distinct thermodynamic costs (see details in Appendix \ref{App:Thermo}), which we assess by comparing the total entropy produced by resetting during the transition time $\tau_{\rm SR}$, to the total entropy produced by the standard yet irreversible potential quench.
We analyze here in details the various entropic contributions in both cases.

\begin{figure}[htb!]
    \centering
    \includegraphics[width=0.9\linewidth]{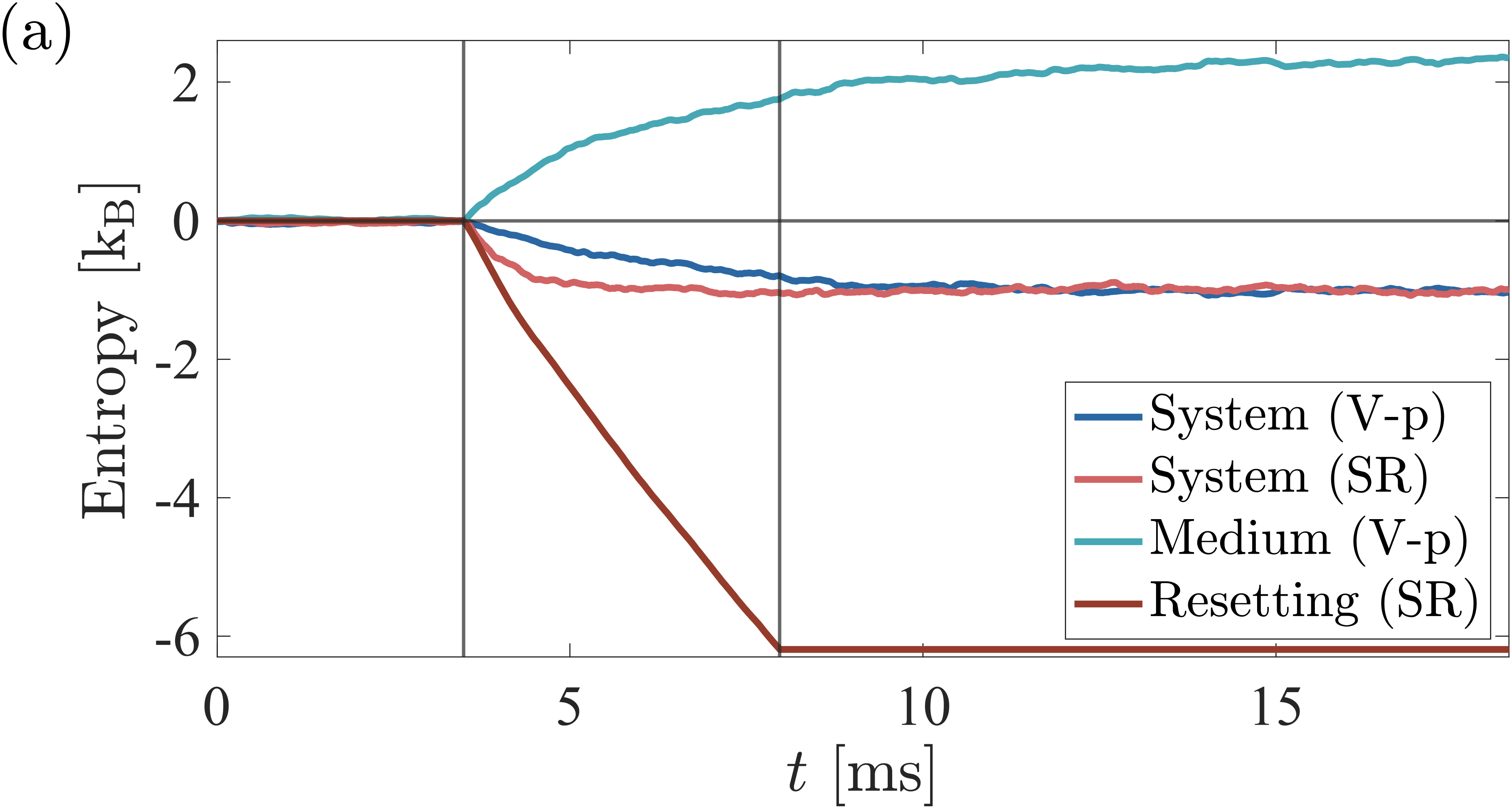}\\
    \includegraphics[width=0.9\linewidth]{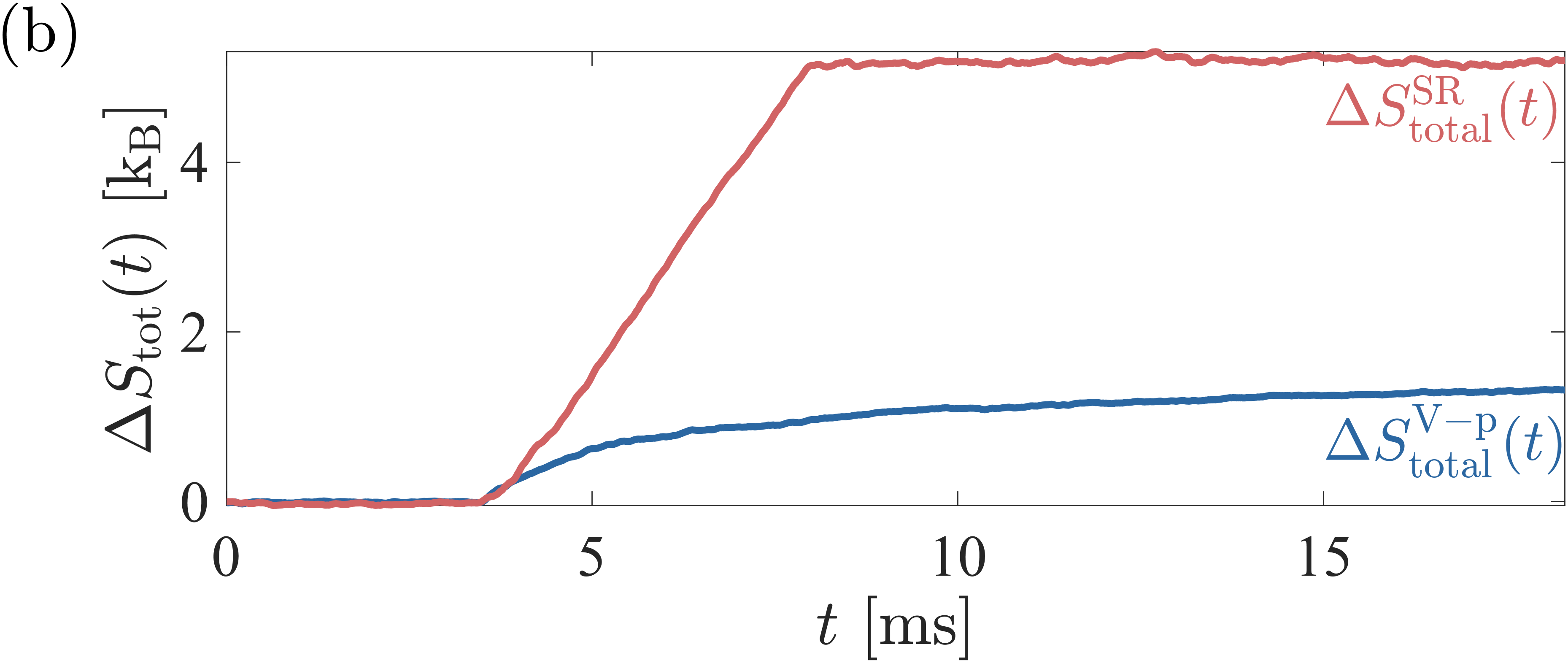}\\
    \includegraphics[width=0.9\linewidth]{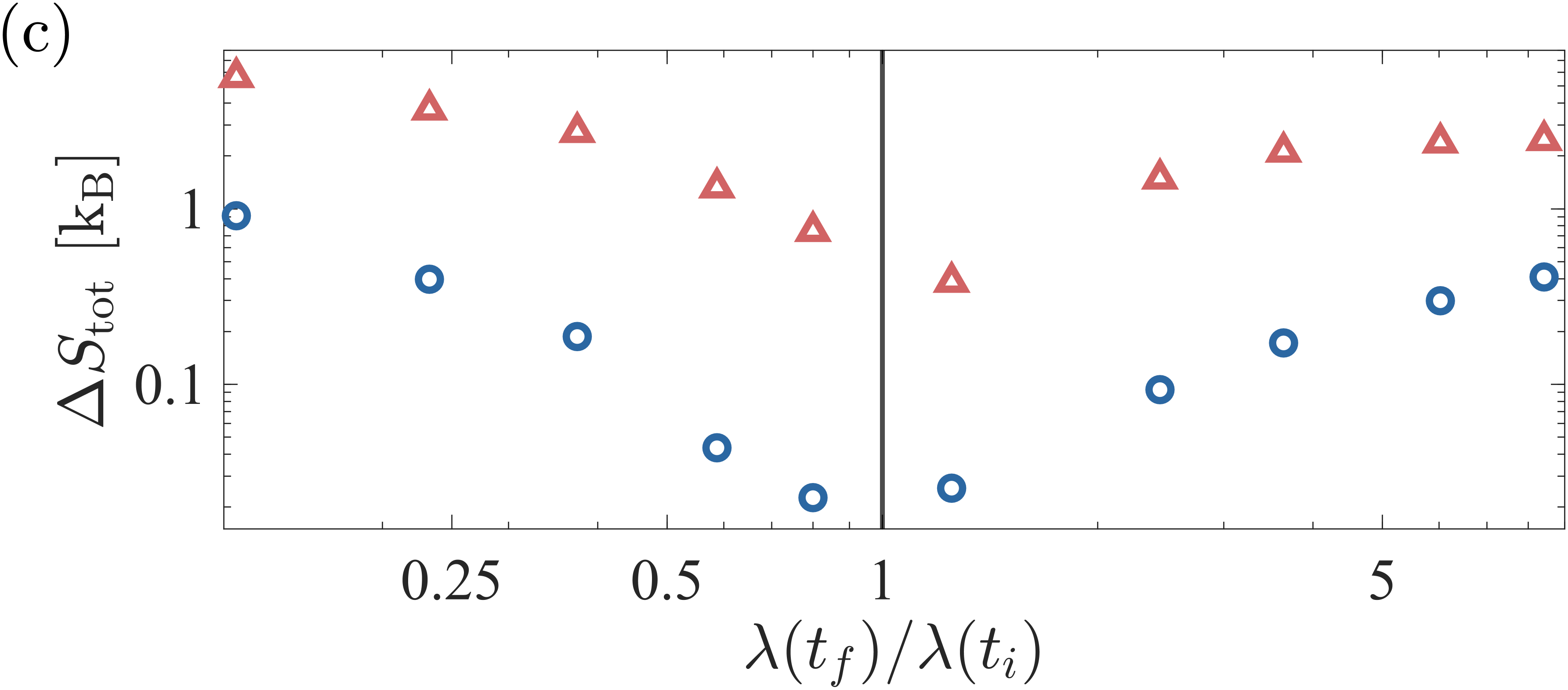}
    \caption{
    (a) Dissipated entropy during the transient for the V-shaped potential (vp) quench (system entropy as deep blue line, medium entropy as light blue line) as well as for the SR-based accelerated protocol (system entropy as red solid line and resetting entropy as brown-red line). In both case the system entropy is the same, as expected since initial and final distributions are equal, only the time-evolution differs.
    (b) Time-dependant total entropy difference $\Delta S(t) = S(t) - S(0)$ for SR (dashed red line) for potential quench (dashed blue line).
    (c) Total entropy as a function of the ratio of states as in Fig.~\ref{fig:KlAndTimes} both for potential quench (blue circles) and SR-based protocol (red triangles).
    }
    \label{fig:Entropy}
\end{figure}

The change of steady-state distribution is accompanied by a change of ensemble-averaged \textit{system entropy} \cite{seifert2005, seifert2012} $S_{\rm sys} = -k_B \int \rho(x,t) \ln \left( \rho(x,t)\right) dx$ (integrals span all $x$).
The total contribution of the state function $\Delta S_{\rm sys}$ along the protocol only depends on initial and final PDFs.
It is the same whether we use a potential quench or resetting.
The system entropy is shown in Fig.~\ref{fig:Entropy}(a) both for a potential quench (deep blue line) and for a resetting-based protocol (red line).
As expected, they converge to the same value which is $\rm{-1~k_B}$ for this choice of parameter (i.e. $\lambda(t_f) = 8 \lambda(t_i)$ as above).
For the potential quench, the heat dissipated in the bath yields a \textit{medium entropy} $S_{\rm m} =  Q / T$ which is calculated by identifying the heat exchanged from the Langevin equation $Q = -\langle \int b(t) \rm{sgn}(x(t)) \dot x(t) dt \rangle$ (with Stratonovich's convention \cite{seifert2012}).
The derivation of the stochastic heat, together with the other entropic contributions, are detailed in Appendix \ref{App:Thermo}.
The heat production for the same potential quench is $Q\approx 2.4~\rm{[k_B T]}$ as shown in Fig.~\ref{fig:Entropy}(a) (turquoise line).
Here, $Q > 0$ showing that during a potential quench, heat is dissipated in the bath.
Since $\dot{Q}$ vanishes in the absence of an external potential, it does not contribute to the SR-based protocol.

Conversely, for the SR-based protocol, the erasure of information during each resetting event yields a \textit{resetting entropy}, which can be understood by focusing on a single resetting event.
Just before resetting the system, the trajectory has diffused until stochastic position $x(t)$ under the action of the thermal bath.
Doing so, it acquired a finite quantity of stochastic system entropy $s(t) = -k_B \ln (\rho(x(t), t))$ where the probability distribution of the system at time $t$ is evaluated at the stochastic position $x(t)$ \cite{seifert2005}.
Immediately after the position is reset to $x(t^+) = 0$  the stochastic system entropy reads $s(t^+) = -k_B \ln (\rho(0, t))$, where we consider that under instantaneous resetting, $\rho(x, t^+) = \rho(x, t)$.
The net difference in system entropy therefore reads $\delta s(t) = s(t^+) - s(t) = k_B \ln(\rho(x(t), t) / \rho(0, t))$.
Resetting events are occurring at a rate $\lambda$ and hence, the average production of resetting entropy is computed as $\dot{S}_{\rm rst} = k_B\lambda \int \rho(x,t) \ln \left( \rho(x,t)/\rho(0,t)\right) dx$  \cite{fuchs_stochastic_2016,pal2017integral, gupta2020work, pal2021thermodynamic, mori_entropy_2023}. 
The total entropy contribution $\Delta S_{\rm rst}$ over the duration $\tau_{SR}$ of our protocol is given by the integral $\Delta S_{\rm rst} = \int_0^{\tau_{SR}} \dot{S}_{\rm rst}(t) dt$.
This is shown on Fig.~\ref{fig:Entropy}(a) (dark red line), yielding a net entropy $\Delta S_{\rm rst} \approx -6.2~\rm{[k_B]}$.
$\Delta S_{\rm rst} <0$ shows the reduction of uncertainty induced by resetting with respect to free diffusion \cite{fuchs_stochastic_2016}.

We can therefore compare the total entropic cost of the SR protocol $\Delta S^{\rm SR}_{\rm tot} = \Delta S_{\rm sys} + \Delta S_{\rm rst} \approx 5.2~\rm{[k_B]}$ to the total entropy produced by the potential quench $\Delta S^{\rm V-p}_{\rm tot} = \Delta S_{\rm sys} + \Delta S_{\rm m} \approx 1.4~\rm{[k_B]}$.
As seen in Fig.~\ref{fig:Entropy}(b), the acceleration obtained by SR comes at the cost of larger entropy production.

For the potential quench, the system transients between two equilibrium states due to the modification of a control parameter \cite{Sekimoto1998, seifert2012}.
The exchanged work $W$ and heat $Q$ obey the first law of thermodynamics and are bounded from below by the free-energy difference between both states $\Delta F = -k_B\log[b(t_f) / b(t_i)]$.
The total entropy production corresponds to the irreversible work $(W - \Delta F)/k_BT \geq 0$, where equality holds for reversible transformations, here, a quasi-static variation of the control parameter $b(t)$. In contrast, an instantaneous quench, as studied here, is inherently irreversible, and this irreversibility is well-captured by the measure of $\Delta S_{\rm tot}$.

For the SR-based protocol, presented in Fig.~\ref{fig:Protocol}, the transition still connects the same equilibrium states, and the free-energy difference $\Delta F$ is not modified.
However, during the transition, the system is driven by resetting, accompanied by constant entropy production when $\lambda \neq 0$.
In contrast to the potential sequence, a quasistatic increase of $\lambda$ would incur a high entropic cost (infinite for true quasistatic protocol) and thus cannot constitute a reversible transition anymore.
Hence, the protocol minimizing total entropy production under SR should have a finite duration. The explicit derivation of such a protocol is beyond the scope of this study. 

While we do not know the optimal protocol to transition between v-shaped potentials and its cost, we can compare the entropic cost of our protocols to known optimized procedures for harmonic traps with the same 8-fold increase of control parameter \cite{Rosales2020}.
In a harmonic potential the total entropy production for a step-like change of stiffness is computed from the dissipated work as $\Delta S_{\rm tot}^{\rm HO, step} = \frac{W_{\rm diss}}{T} = 2.2 ~ \rm{k_B}$.
Using an optimal protocol to decrease the relaxation time by the same factor $3.5$ as obtained here, would require $\Delta S_{\rm tot}^{\rm HO, opt} = 10.6 ~ \rm{k_B}$.
Both numbers are of the same order of magnitude as the entropic cost of the protocol proposed here, showing that the SR-induced dissipation does not exceed useful known protocols.

Experimental implementations of SR of colloidal particles in optical traps \cite{tal2020, besga2020, goerlich2023} necessarily consist of finite-time resetting events \cite{gupta2020stochastic, bodrova2020resetting, gupta2021resetting, olsen2023thermodynamic}.
One example of such a resetting procedure is constant-time resetting \cite{besga2020, goerlich2023}, where an external stiff potential is applied for a finite time to ensure the particle returns to the origin.
For our proposed method to be advantageous, the total sum of individual resetting times should not surpass the difference between $\tau_r$ and $\tau_{\text{SR}}$.
This condition establishes a maximum temporal cost for each resetting event (see Appendix \ref{App:Thermo} for specifics) well within experimental capabilities. Taking into account the range of relaxation times shown in Fig.~\ref{fig:KlAndTimes}, the maximal temporal costs range is $\approx 1-100$ ms. This is notably larger than the fraction-of-millisecond duration realized in previous experiments \cite{goerlich2023}.

\textit{Conclusions} - Our results show that SR can be used to accelerate transitions between two equilibrium states under different v-shaped potentials.
This method uses a non-deterministic driving scheme, unlike standard SST deterministic protocols \cite{guery-odelin_driving_2023}. Moreover, it naturally applies to transitions between non-equilibrium steady states and provides a simple acceleration scheme between non-harmonic potentials. The thermodynamic cost of SR-induced acceleration is equivalent to that of the previously suggested deterministic SSTs for similar accelerations, even when considering the addition of the finite duration of the resetting process.

The general mechanism to accelerate transitions between states is manipulating the probability currents' evolution. This 
drives the system far from equilibrium and enables circumventing the limitations imposed by equilibrium conditions. Macroscopically, this was previously achieved by dynamic modification of the potential landscape. Our approach induces these currents by modifying the microscopic dynamics. However, other ways to control the probability current, such as transiently inducing self-propulsion, may prove beneficial and extend the available tools for SST.

Our method extends beyond the specific transitions observed between v-shaped potentials studied in this context. It has the potential for generalization across diverse potentials by utilizing more sophisticated SR protocols. For instance, the application of partial resetting \cite{tal2022}, involving resetting to a fraction of the distance to the origin, allows for the fine-tuning of the steady-state probability distribution of a diffusing particle. This adjustment spans the entire spectrum between Gaussian and Laplace distributions, facilitating transitions among families of potentials. Additionally, implementing state-dependent resetting \cite{Roldan2017} further broadens the repertoire of manipulable states. However, both these strategies necessitate feedback mechanisms based on measurements for their operation.

The acceleration mechanism revealed in this study likely underlies the modification of efficiency induced by SR when implemented in a stochastic heat engine \cite{lahiri_can_2023}. Thus, gaining a deeper understanding of the interaction between acceleration and the thermodynamic cost of SR, specifically in SST, is crucial for leveraging its potential in pioneering micromachine applications \cite{blickle2012realization, martinez2016brownian, martinez2017colloidal}.\\

\section*{Acknowledgments}
 We are grateful to Shlomi Reuveni for illuminating conversations. Y.R. acknowledges support from the Israel Science Foundation (grants No. 385/21). Y.R and R.G. acknowledge support from the European Research Council (ERC) under the European Union’s Horizon 2020 research and innovation programme (Grant agreement No. 101002392). R.G. acknowledges support from the Mark Ratner Institute for Single Molecule Chemistry at Tel Aviv University.
  
 \appendix

\section{\label{App:Thermo} Stochastic energetics}

\textbf{Potential quench - }
For the transition between two equilibrium states in v-shaped potential $V(x, t) = b(t) |x|$, the system obeys the following Langevin equation
\begin{equation}
    \dot{x}_t = \frac{-b(t)}{\gamma} \rm{sgn}(x_t) + \sqrt{2 D} \xi_t
\end{equation}
where $D = k_B T / \gamma$ with $T$ the bath temperature, $k_B$ Boltzmann's constant and  $\gamma$ Stokes viscous drag.
$\rm{sgn}(x_t)$ is the sign function applied to the stochastic variable $x_t$ and $\xi_t$ a Gaussian white noise.
Following the approach developed by Ken Sekimoto \cite{Sekimoto1998}, the Langevin force balance equation can be turned into an energy balance by multiplying each side by the spatial increment $dx_t$ undergone by the system within a time $dt$
\begin{equation}
    \left( \gamma \dot{x}_t - \sqrt{2 k_B T \gamma } \xi_t \right) \circ dx_t = -b(t) \rm{sgn}(x_t) \circ dx_t 
\end{equation}
where $\circ$ denotes Stratonovich convention \cite{seifert2012}. The left hand side corresponds to the energy exchanged through the action of the solvent molecules, it is associated with the heat dissipated as
\begin{equation}
    \delta q = \left( \gamma \dot{x}_t - \sqrt{2 k_B T \gamma} \xi_t \right) \circ dx_t.
\end{equation}
The right hand side corresponds to the derivative of the potential with respect to the variable $x$. Since here the coefficient $b(t)$ can vary in time, the total derivative of the potential, which is the change of internal energy of the system, reads
\begin{equation}
    du = d V[x, b(t)] = \frac{\partial V}{\partial x} \circ dx + \frac{\partial V}{\partial b(t)} db(t).
\end{equation}
The second term in the internal energy difference, $\frac{\partial V}{\partial b} db$ corresponds to the energy exchanged due to the modification of an externally controlled parameter. It is associated with work
\begin{equation}
    \delta w = \frac{\partial V}{\partial b} \frac{d b}{dt} dt = |x_t| \dot{b}(t) dt.
\end{equation}

\begin{figure}[htb!]
    \centering
    \includegraphics[width=1\linewidth]{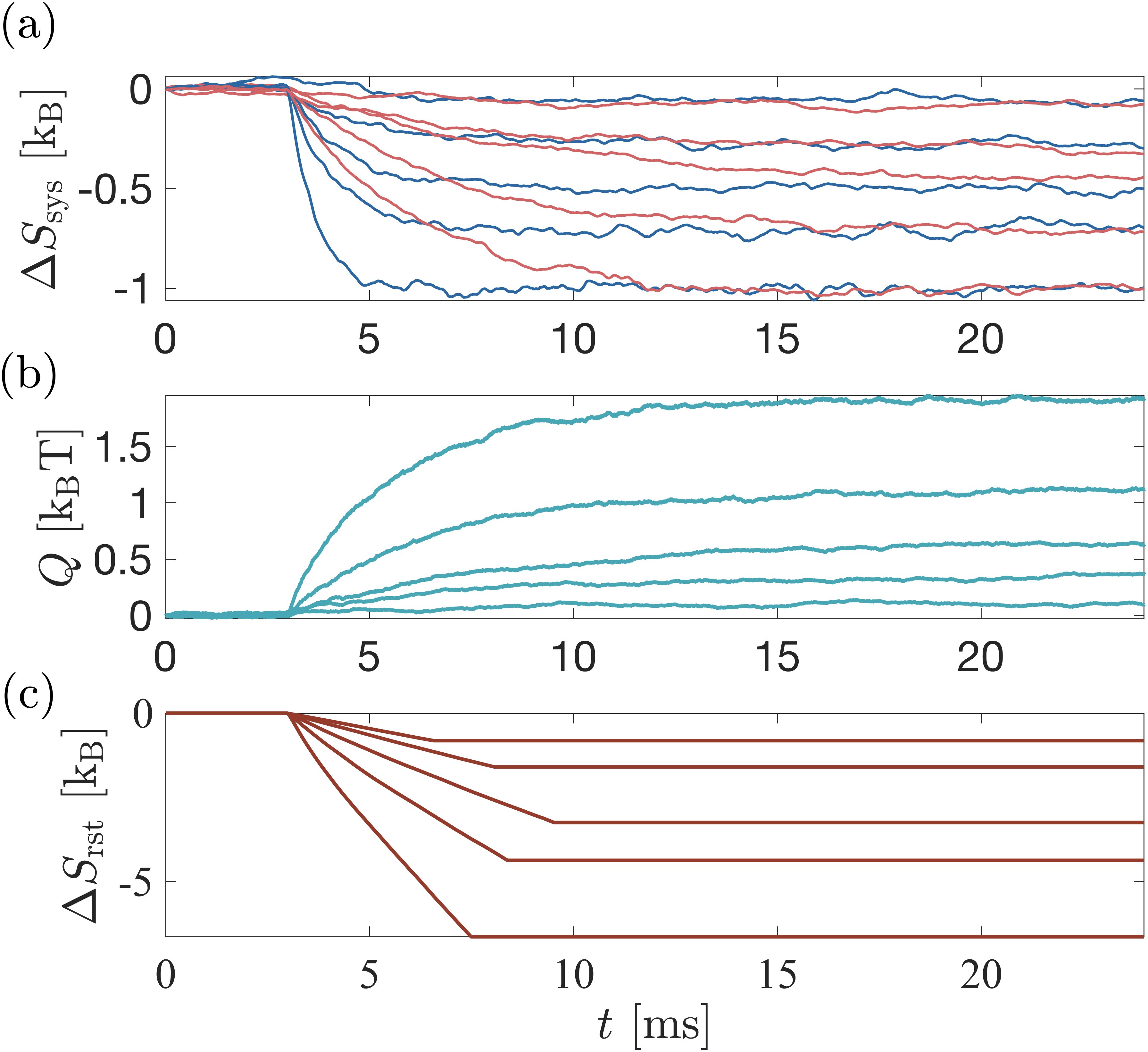}
    \caption{ Entropy produced during a transition with $\lambda(t_f)/\lambda(t_i)$ spanning from $1/1.25$ to $1/8$. For all cases, the entropy produced increases as the difference between states increases.
	(a) system entropy for both potential quench (blue line) and SR-induced transition (red line)
    (b) Heat dissipation for the potential quench
    (c) Integrated resetting entropy production rate during the SR-induced transition. The non-monotonic evolution of the associated SR-induced relaxation time $\tau_{\rm SR}$ is visible as the second inflection point in the various curves.
    }
    \label{fig:AllEntropyDown}
\end{figure}

The first law of thermodynamics then reads $du = \delta w - \delta q$ at the level of each trajectory, where each energetic quantity is a stochastic variable.
By the first law, heat can also be evaluated as
\begin{equation}
    \delta q = \delta w - du = -b(t) \rm{sgn}(x_t) \circ dx_t
\end{equation}
which can be easily measured numerically or experimentally on stochastic trajectories $x_t$.
When an ensemble of trajectories undergo the same quench of potential $b(t)$ as studied in the main text of the paper, the stochastic heat $\delta q$ can be evaluated on each trajectory and the ensemble-averaged heat at a time $t$ is the accumulation of ensemble-averaged increments as $Q^{\rm vp}(t) = \int_0^t \langle \delta q_t \rangle$, where ``vp'' denotes the V-shaped potential quench thermodynamics.
This quantity is plotted on Fig. \ref{fig:Entropy} (a) in the main text as a blue line for the same 8-fold quench of $\lambda$, resulting in a total $Q^{\rm vp}(t_f) = 2.4 ~ \rm{k_B T}$ of dissipated heat.
It is also plotted on Fig.~\ref{fig:AllEntropyDown} (b) for various ratio of $\lambda(t_f)/\lambda(t_i)$.

The dissipated heat can also be expressed as an entropy dissipated in the medium $S^{\rm vp}_{\rm m} = Q^{\rm vp} / T$. However, as the potential is quenched the entropy associated to the available configuration space also changes. This \textit{system entropy} takes here the form of Gibbs entropy
\begin{equation}
    S^{\rm vp}_{\rm sys} = -k_B \int_{-\infty}^{+\infty} \rho(x,t) \rm{ln}[\rho(x,t)] dx
\end{equation}
which is the ensemble averaged trajectory dependent entropy $-k_B \rm{ln}[\rho(x_t, t)]$ \cite{seifert2005}.
This stochastic entropy can be evaluated through time on each trajectory undergoing a potential quench and $S^{\rm vp}_{\rm sys}$ recovered through ensemble-average.
It is plotted as a blue line on Fig.~\ref{fig:Entropy} and on Fig.~\ref{fig:AllEntropyDown} (a) for more cases.
$S^{\rm vp}_{\rm sys}$ is a state function which only depends on the initial and final values of the probability densities $\rho(x, t_i)$ and $\rho(x, t_f)$.
The total entropy production during the potential quench is the sum of both \textit{medium} and \textit{system} entropies as
\begin{equation}
    S_{\rm tot}^{\rm vp} = \frac{Q^{\rm vp}}{T} + S^{\rm vp}_{\rm sys} \geq 0.
\end{equation}
The thermodynamic cost associated with the potential quench can be evaluated either through its energetic footprint via the first law or through its entropic cost via the second law.\\

\textbf{Resetting rate quench - }
In the case of the stochastic resetting work $W^{\rm SR}$ and heat production rates can also be evaluated as $\dot{W}^{\rm SR} = -\lambda \langle V(x_t, t) \rangle$ and $\dot{Q}^{\rm SR} = \int F(x, t) j(x,t) dx$ where $F(x,t)$ is the force at time $t$ and $j(x,t)$ the probability current, consequence of the NESS nature of SR \cite{fuchs_stochastic_2016}.
Under constant potential $V(x)$, the first law then simply reads $\dot{W}^{\rm SR} + \dot{Q}^{\rm SR} = 0$.
Importantly, both quantities vanish in the absence of an external confining potential $V(x,t)$.

The entropy associated with SR breaks into three distinct contributions
\begin{equation}
    \dot{S}_{\rm tot}^{\rm SR} = \frac{\dot Q^{\rm SR}}{T} + \dot{S}_{\rm sys}^{\rm SR} - \dot S_{\rm rst}^{\rm SR} \geq 0
\end{equation}
where the first and second term keeps the same interpretation as above. The heat production being zero, the first term vanishes in the absence of potential. The system entropy being a state function which only depends of the initial and final densities, its total production along a $\lambda$-quench will be the same as in the potential quench (the initial and final densities being the same in both cases).
This is verified on Fig.~\ref{fig:Entropy} (a) in the main text where in the long-time limit we see $S_{\rm sys}^{\rm SR}(t_f) = S_{\rm sys}^{\rm vp}(t_f) = 1 ~ \rm{k_B}$ (albeit in a shorter time for the SR-base protocol).
It is also plotted on Fig.~\ref{fig:AllEntropyDown} (c) for more cases.
The third term in the second law for SR is the so-called \textit{resetting entropy} production rate \cite{fuchs_stochastic_2016, mori_entropy_2023}.
It is associated with the constant erasure of information at play during a SR process, each resetting event erasing the information stored in the stochastic position $x_t$ \cite{goerlich2023}.
It is never zero, denoting the NESS nature of SR which constantly produced entropy.
In initial and final NESSes, the production rate is constant, during the transient it becomes a time-dependent function.

In the protocol proposed in the main text, SR is used only during the transient time $\tau_{\rm SR}$ to bring the system in its new equilibrium state in the final external potential.
The entropic cost of such protocol can therefore be evaluated as
\begin{equation}
    \Delta S^{\rm SR} = \int_0^{\tau_{\rm SR}} \dot S_{\rm tot}^{\rm SR}(t) dt = \Delta S_{\rm sys}^{\rm SR} - \int_0^{\tau{\rm SR}} \dot S_{\rm rst}^{\rm SR} dt.
\end{equation}
where we recall that $\Delta S_{\rm sys}^{\rm SR} = \Delta S_{\rm sys}^{\rm vp}$ has the same contribution in the case of a potential quench.
The cost, i.e. the cumulative integral of this quantity is plotted on Fig.~\ref{fig:Entropy} (a) in the main text as well as on Fig.~\ref{fig:AllEntropyDown} (c) for more cases.
Finally, the measure of the entropic cost of the proposed accelerated protocol proceeds in comparing minus the integrated resetting entropy production rate $-\int_0^{\tau{\rm SR}} \dot S_{\rm rst}^{\rm SR} dt = 6.2 ~ \rm{k_B}$ to the heat dissipated by the standard potential quench $\Delta Q^{\rm vp}/T =(Q^{\rm vp}(t_f) - Q^{\rm vp}(t_i)) / T = 2.4 ~ \rm{k_B}$.\\

\textbf{Comparison with known optimal techniques in harmonic potentials - }
In the case of harmonic external potentials $V(x) = \kappa x^2 / 2$, SST methods exist \cite{martinez_engineered_2016} and the full characterization of the energetics associated allowed the derivation of optimal protocols \cite{Rosales2020} were the dissipated work (and heat) is minimized for a given acceleration with respect to thermal relaxation.
Here we propose to compare the entropic cost of the SR-induced acceleration to the entropy generated by an optimal protocol imposing the same 8-fold increase of the control parameter and leading to the same 3.25 acceleration.
The physical meaning of such comparison is of course limited since the optimal protocol applies on a harmonic potential and while we work with linear v-shaped potential.
However, the fact that we obtain a very similar cost is a promising result for the generalization of SR-induced acceleration.\\

Two states defined respectively by $\kappa_i$ and $\kappa_f$ are characterized by a free-energy difference $\Delta F = k_B T \rm{ln}\sqrt{\kappa_f/\kappa_i}$.
For a step-like change of stiffness from $\kappa_i$ to $\kappa_f$ the work reads $\Delta W^{\rm ho, step} = k_B T (\kappa_f - \kappa_i)/(2\kappa_i)$.
For an optimal protocol imposing an n-fold acceleration $\Delta t \equiv \tau_{r} / n$, the work reads $\Delta W^{\rm ho, opt} = \gamma (\sqrt{k_B T / \kappa_i} - \sqrt{k_B T / \kappa_f})^2/(2 \Delta t) + \Delta F$ as detailed in \cite{Rosales2020}.
The total entropic cost in both cases reads $\Delta S_{\rm tot}^{\rm ho} = (W^{\rm ho} - \Delta F)/ T$ which corresponds to the dissipated work divided by the temperature.
Feeding in $\kappa_f = 8 \times \kappa_i$ and $\Delta t = \tau_{r} / 3.5$ to stick to the conditions of the protocol studied here, on obtains $\Delta S_{\rm tot}^{\rm ho, step} = 2.16 ~ \rm{k_B}$ and $\Delta S_{\rm tot}^{\rm ho, opt} = 10.64 ~ \rm{k_B}$.
Those numbers are close to the obtained results for the v-shaped step-like potential quench $\Delta S_{\rm tot}^{\rm eq} = 2.96 ~ \rm{k_B}$ and for the SR-induced protocol $\Delta S_{\rm tot}^{\rm SR} = 6.3 ~ \rm{k_B}$.

\section{Compression and expansion}
\label{App:detailsMoments}

We detail here the differences between expansion (decrease of $\lambda$ or $b$) and compression (increase of $\lambda$ or $b$).

\begin{figure}[htb!]
    \centering
    \includegraphics[width=0.9\linewidth]{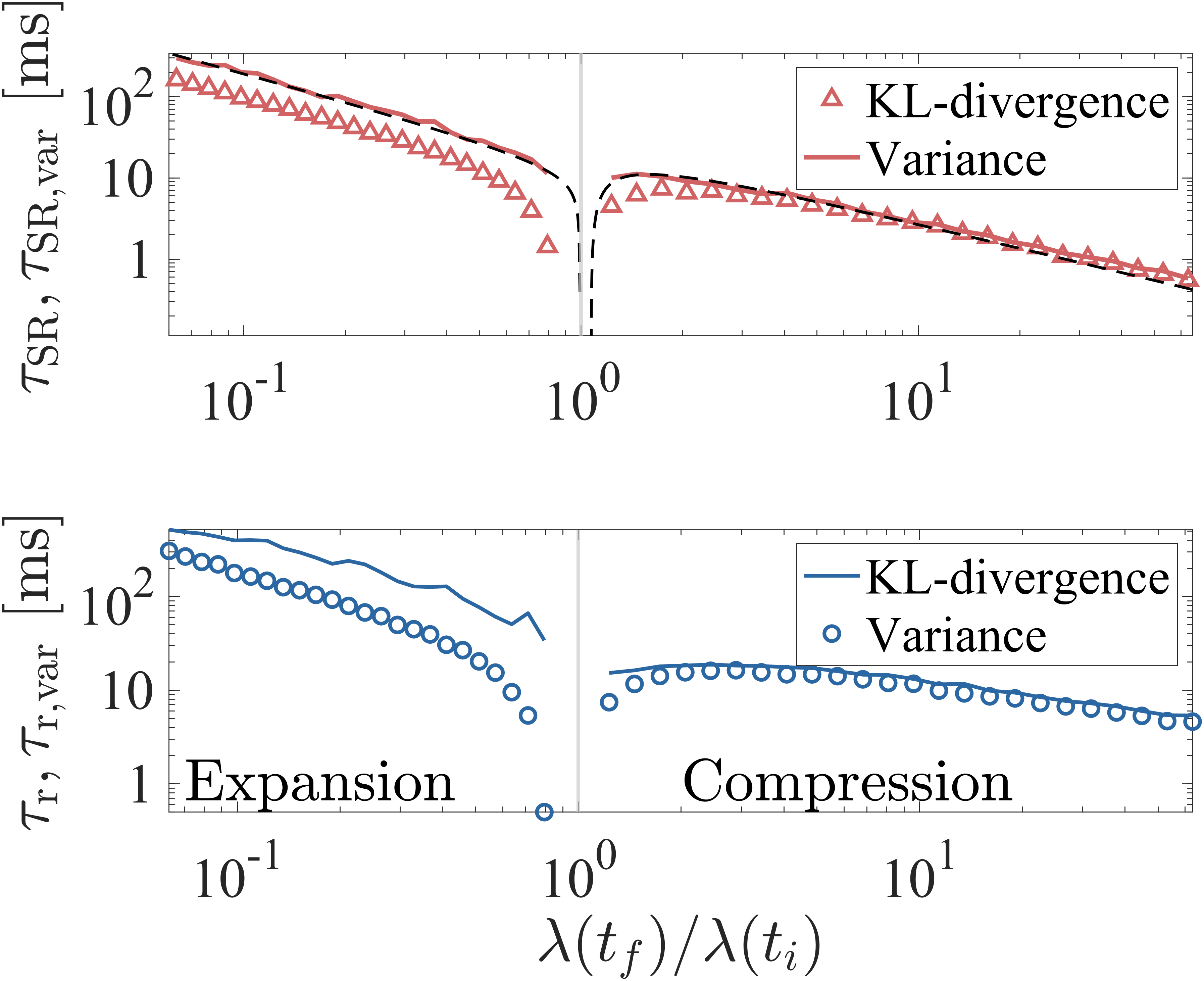}
    \caption{
	Relaxation times measured for the KL-divergence (symbols) and on the variance (solid lines). In both case, it corresponds to the time of the first point of the respective quantity to reach the mean final value.
    On the top graph for a SR-based protocol (with result Eq.~(\ref{Eq:FullRelaxTime}) as black dashed line) and on the lower panel for a potential quench.
    }
    \label{fig:detailsRelaxTimes}
\end{figure}

The relaxation time measured on the KL-divergence is the time corresponding to the first point in the KL-divergence to reach its final average value i.e. 1.
We refer to this time as the \textit{full relaxation time}.
The full relaxation time can be evaluated for variance as well, and the result is shown Fig~\ref{fig:detailsRelaxTimes}.
It should not be confused with the exponential characteristic time of the second moment $\tau_{\rm exp}$, shown in the inset of Fig~\ref{fig:KlAndTimes}.
The analytical expression Eq.~(\ref{eq:var}) of the variance under SR allows us to obtain an exact expression for the time needed for the variance to reach a value arbitrarily close to the final steady-state value.
More precisely, we search for the time $\tau_{\rm{SR, var}}$ such that $S(\tau_{\rm{SR, var}}) = 2 D / \lambda(t_f) + \epsilon$ where $\epsilon$ can be arbitrarily small.
First, the variance equation is expressed in dimensionless units of space $\bar x = \sqrt{\frac{\lambda(t_f)}{2 D}}x$ and time $\bar t = \lambda_f t$ as
\begin{equation}
    \bar S(\bar t) = \langle \bar{x}^2(~\bar{t}~)\rangle =  (\theta - 1) e^{-\bar t} + 1,
\end{equation}
where $\theta = \lambda(t_i) / \lambda(t_f)$.
The full relaxation condition becomes $\bar S(\bar t) = 1 + \epsilon$ and we obtain,
\begin{equation}
    \bar \tau_{\rm{SR, var}} = -\ln(\epsilon) + \ln(|1 - \theta|),
    \label{Eq:FullRelaxTime}
\end{equation}
where the absolute values apply to both compression and expansion.
The tolerance parameter $\epsilon$ is arbitrary and the relation between the relaxation time and $\epsilon$ depends on the simulation statistics.
We therefore fit the variance relaxation time shown in Fig.~\ref{fig:detailsRelaxTimes} (black dashed line) with $\bar \tau_{\rm{SR, var}} = \alpha ( \beta + \ln(|1 - \theta|))$ with $\alpha$ and $\beta$ as fitting parameters.
The same function Eq.~\ref{Eq:FullRelaxTime} is used to fit the full relaxation of the KL-divergence in Fig~\ref{fig:KlAndTimes}, showing that both variance and KL-divergence relax following similar time-scales.

\begin{figure}[htb!]
    \centering
    \includegraphics[width=1\linewidth]{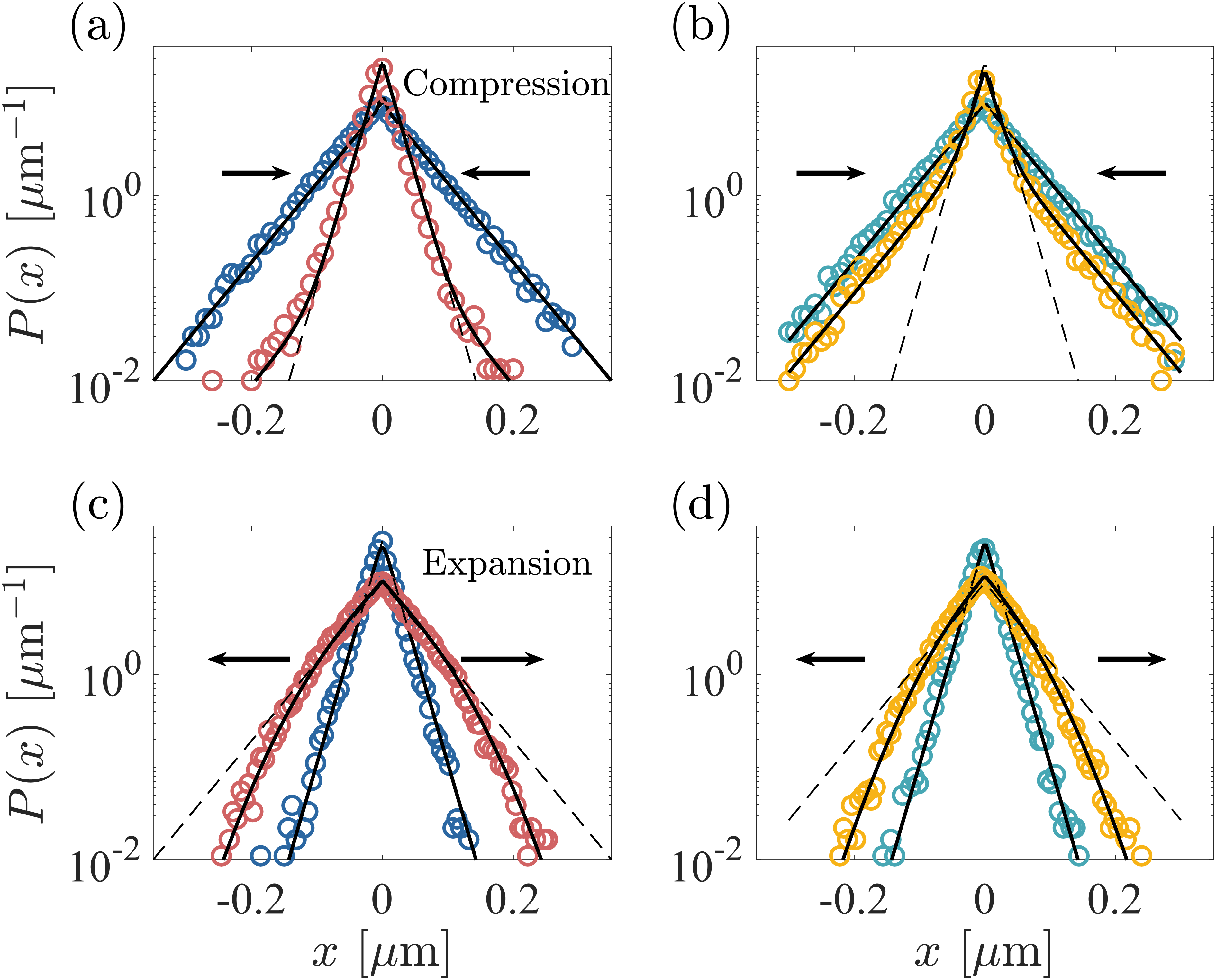}\\
    \caption{Numerically measured histogram and associated exact result (black solid lines).
	(a) SR-based compression from $\lambda(t_i) = 200$ to $\lambda(t_f) = 1600$ Hz, we represent both the initial state  (blue) and a transient state after $t = 2.2$ ms (red).
    (b) V-shaped potential compression between similar states, both initial state (turquoise) and transient (yellow)
    (c) SR-based expansion from $\lambda(t_i) = 1600$ to $\lambda(t_f) = 200$ Hz, both initial state and transient after $t = 3$ ms.
    (d) V-shaped potential expansion between similar states.
    }
    \label{fig:HistCompExp}
\end{figure}

Interestingly,  the response to compression and to expansion are very different.
For compression, both for SR-based transformation (Fig.~\ref{fig:HistCompExp}(a)) and potential quench (Fig.~\ref{fig:HistCompExp}(b)), the distribution takes a bi-Laplace shape, well captured by both the exact result $\rho_{\rm exact}(x,t)$ and our empirical model $f(x,t)$.
The front separating both regions travels from the center towards the tails.

For expansion, both for SR-based transformation (Fig.~\ref{fig:HistCompExp}(c)) and potential expansion (Fig.~\ref{fig:HistCompExp}(d)), the distribution possesses a Laplace core, but its tails have a Gaussian character.
Again, the front separating both regions travels from the center towards the tails.
For the SR-based expansion, the system is prepared during $t<0$ in a state corresponding to a high resetting rate and is evolving during $t>0$ with a lower resetting rate.
Namely, the resetting events become abruptly less frequent.
We, therefore, interpret both regions as follows: (1) the core of the distribution corresponds to trajectories that already have undergone resetting with the new, lower rate, and (2) the Gaussian tails correspond to trajectories that are still freely diffusing.
The tails are not purely Gaussian because these transiently freely diffusing trajectories are initially distributed along a Laplace profile.
For the potential expansion, the sudden decrease of the potential slope corresponds effectively to an increase of available volume.
Here again, trajectories will transiently experience a free expansion in the new volume, bearing a Gaussian profile.

On the different panels of Fig.~\ref{fig:HistCompExp}, we show measured histograms in the initial state and during the transient, illustrating these different cases.
The exact time-dependent distribution (black solid line) always matches the measured histograms.

\section{Transient probability density for SR}
\label{App:PDF}
In this appendix we derive analytically the time-dependant PDF for an SR process undergoing a sudden change of resetting rate $\lambda$.

We consider a particle diffusing under stochastic resetting with resetting rate $\lambda_i$ and diffusion coefficient $D$ which already reached its non-equilibrium steady state, given by the well-known Laplace distribution $P_s(x)=\sqrt{\frac{\lambda_i}{4D}}e^{-\sqrt{\lambda_i/D}|x|}$ \cite{evans_diffusion_2011}. At time $t=0$, the resetting rate is changed to $\lambda_f$. We look for the probability $f_s(x,t)$ of finding the particle at position $x$, given time $t$. We will define a Green's function $g_s(x,t|y)$, which is the probability of finding the particle at position $x$, given time $t$, and given that the particle was at position $y$ at $t=0$. $f_s(x,t)$ is given by

\begin{align}\label{Eq: resetting_prop_int}
    f_s(x,t) &=\int_{-\infty}^\infty g_s(x,t|y)P_s(y)\, dy\Rightarrow \Tilde{f}_s(x,s)\\ &= \int_{-\infty}^\infty \Tilde{g}_s(x,s|y)P_s(y)\, dy,
\end{align}
where $\Tilde{f}_s(x,s)$ and $\Tilde{g}_s(x,s|y)$ are the Laplace transforms of the propagator $f_s(x,t)$ and of the Green's function $g_s(x,t|y)$ evaluated at Laplace variable $s$, respectively. The Laplace transform of the Green's function for stochastic resetting was calculated before using renewal theory and is given by $\Tilde{g}_s(x,s|y)=\Tilde{G}_0(x,\lambda_f+s|y)+\frac{\lambda_f}{s}\Tilde{G}_0(x,\lambda_f+s|0)$, where $\Tilde{G}_0(x,s|y)$ is the Laplace transform of the free diffusion propagator given by $\Tilde{G}(x,s|y)=\frac{1}{2\sqrt{sD}}e^{-\sqrt{\frac{s}{D}}|x-y|}$ \cite{evans_stochastic_2020}. Plugging everything back to Eq.~(\ref{Eq: resetting_prop_int}) gives
\begin{widetext}
\begin{equation}
    \Tilde{f}_s(x,s)=\frac{\lambda_f}{2\sqrt{(\lambda_f+s)D}}e^{-\sqrt{\frac{\lambda_f+s}{D}}|x|}+\sqrt{\frac{\lambda_i}{s+\lambda_f}}\frac{1}{4D}\int_{-\infty}^\infty e^{-\sqrt{\frac{s+\lambda_f}{D}}|x-y|-\sqrt{\frac{\lambda_i}{D}}|y|}\, dy.
\end{equation}
Solving the integral gives the propagator for the relaxation in Laplace space
\begin{equation}\label{Eq: Resetting Relax Propagator}
    \Tilde{f}_s(x,s)=\frac{1}{2\sqrt{D}(\lambda_i-\lambda_f-s)}\left[\frac{(\lambda_i-\lambda_f)\sqrt{\lambda_f+s}}{s}\exp{\left(-\sqrt{\frac{\lambda_f+s}{D}}|x|\right)}-\sqrt{\lambda_i}\exp{\left(-\sqrt{\frac{\lambda_i}{D}}|x|\right)} \right].
\end{equation}
\end{widetext}

We can see that taking the limit $\lim_{s\to0} s\Tilde{f}_s(x,s)$ reproduces the non-equilibrium steady state with resetting rate $\lambda_f$. One can invert the Laplace transform numerically and obtain the time-dependent PDF,$\rho_{\rm exact}(x,t)$, using Eq.~(\ref{Eq: Resetting Relax Propagator}).
$\rho_{\rm exact}(x,t)$ agrees well with our empirical PDF $f(x,t)$ defined in Eq.~(\ref{eq:TimeDepPDFMain}). Both are plotted together with the numerical histograms in Fig.\ref{fig:HistCompExp}.
Here, like in the case studied in \cite{Majumdar2015} a single time-dependant front $x_0(t)$ splits both domain.
The time-dependence of $x_0(t)$ is the only parameter when fitting the measured histograms to $f(x,t)$ and the time-dependant result is shown in Fig. \ref{fig:ContourProtocol} of the main text.

\begin{figure}[htb!]
    \centering
    \includegraphics[width=0.9\linewidth]{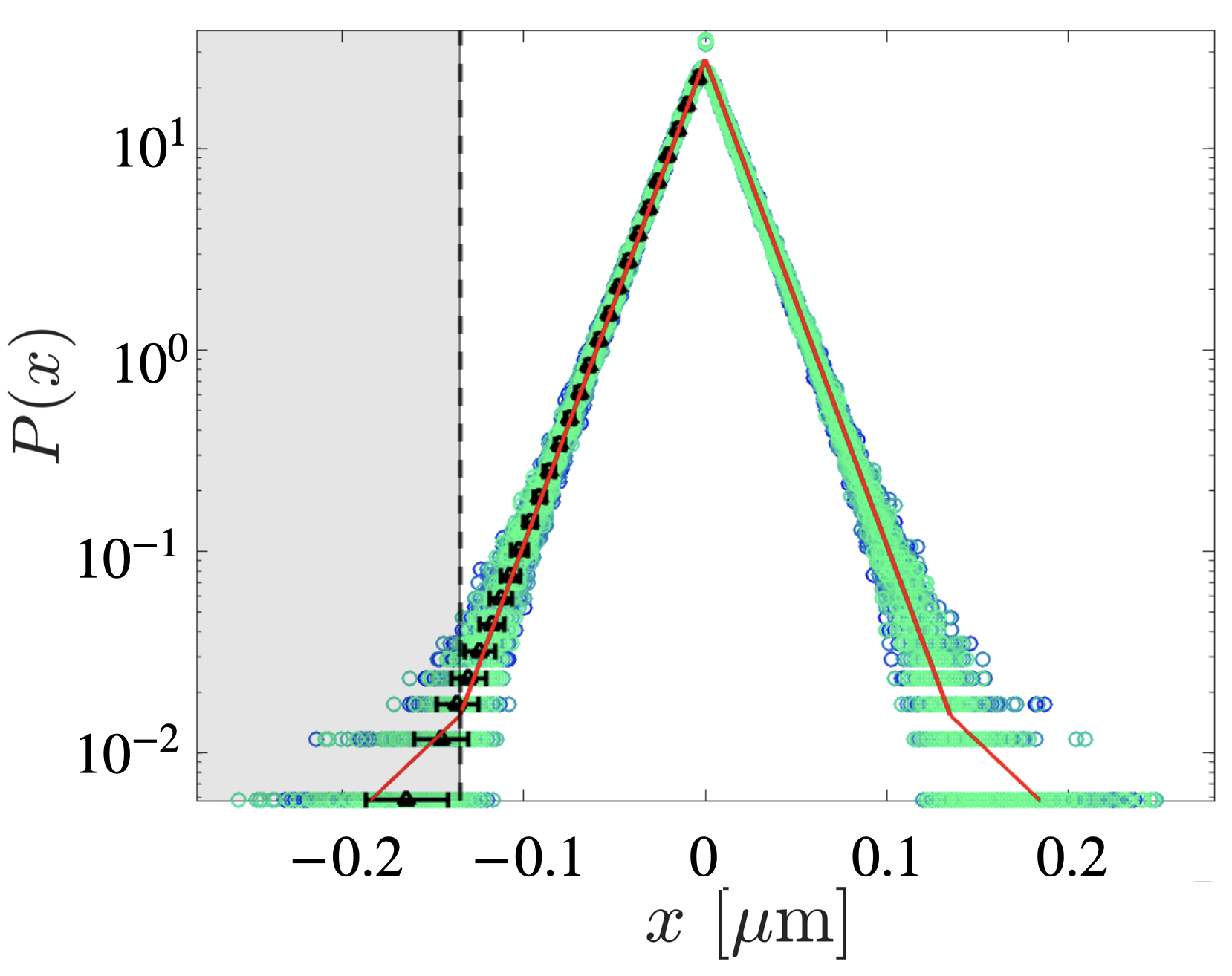}
    \caption{
	Evaluation of the limit on the precision of a numerical evaluation of $x_0(t)$.
    A thousand consecutive measured PDF on an ensemble on $10^5$ trajectories in final steady-state (blue to green), average and standard deviation of the measured position at a given value $P_f(x)$ (black triangle) and transient $f(x,t)$ defined by $x_0(t) = 1.3 \times 10^{-7}~\rm{m}$.
    }
    \label{fig:PrecisionX0}
\end{figure}

Importantly, the numerically measured values of $x_0(t)$ only keeps a physical significance in a defined range.
On Fig. \ref{fig:PrecisionX0} we show the dispersion in numerical histogram's tails.
This will lead to an upper bound on the measurable values of $x_0$
%Indeed, $x_0$ bei,g extracted from the fitting of Eq.~(\ref{eq:TimeDepPDFMain}) on the measured PDFs.
As seen here with the red line, $f(x,t)$ defined by $x_0(t) = 1.3 \times 10^{-7} ~\rm{m}$ falls into the errobars on the estimation of $x$ until the tail of the histogram.
Any larger value of $x_0(t)$ will also fall within the same standard deviation. We therefore use this value of $x_0(t) = 1.3 \times 10^{-7}~\rm{m}$ as an upper bound set by numerical precision (it corresponds to the limit of the $x_0(t)$ plot the main text) for this given statistics.

\section{Relaxation of the Moments in Stochastic Resetting}
\label{App:Variance}

The propagator in Eq.~(\ref{Eq: Resetting Relax Propagator}) is symmetric around the origin. Hence, all the odd moments are zero at any time. We compute the even moments $\langle \Tilde{X}^{2n}(s)\rangle$ using the fact that $\int_{-\infty}^{\infty} x^{2n}e^{-\alpha|x|}\, dx=2\Gamma(2n+1)\alpha^{-2n-1}$, where $\Gamma(y)$ is the gamma function. The even moments are given by,
\begin{equation}
    \langle \Tilde{X}^{2n}(s)\rangle=\frac{\Gamma(2n+1)D^n}{\lambda_i-\lambda_f-s}\left[\frac{\lambda_i-\lambda_f}{s(\lambda_f+s)^n}-\frac{1}{\lambda_i^n}\right].
\end{equation}
We can describe the relaxation by calculating the difference between the time-dependent moment and its final value in the steady state. The time-dependent moment is obtained by inverting the aforementioned Laplace transform.The even moments at the final steady state are given by, $\Gamma(2n+1)(D/\lambda_f)^n$, leading to, 
\begin{align}
    &\langle X^{2n}(t)-X^{2n}_{ss}\rangle = \\
    &\frac{\Gamma(2n+1)D^n}{\Gamma(n)}\left[\frac{\Gamma(n,\lambda_it)}{\lambda_i^n}e^{(\lambda_i-\lambda_f)t}-\frac{\Gamma(n,\lambda_ft)}{\lambda_f^n}\right],
\end{align}
where $\langle X^{2n}_{ss}\rangle$ is the $2n$ moment at the steady state, and $\Gamma(n,x)=\int_{x}^\infty y^{n-1}e^{-y}\, dy$ is the upper incomplete gamma function. In the limit of $t\gg \lambda_i^{-1},\lambda_f^{-1}$ we can approximate the incomplete gamma function by $\Gamma(n,x)\sim x^{n-1}e^{-x}$ to obtain
\begin{equation}
    \langle X^{2n}(t)-X^{2n}_{ss}\rangle\sim\left(\frac{1}{\lambda_i}-\frac{1}{\lambda_f}\right)\frac{\Gamma(2n+1)}{\Gamma(n)}D^nt^{n-1}e^{-\lambda_ft}.
\end{equation}
At long times, the moments decay like a power law multiplied by an exponential with a characteristic rate equal to the new resetting rate. Because $\Gamma(1,x)=e^{-x}$,  the variance decays exponentially for all times as,
\begin{equation}
    \langle X^2(t)-X^2_{ss}\rangle=2D\left(\frac{1}{\lambda_i}-\frac{1}{\lambda_f}\right)e^{-\lambda_ft}.
\end{equation}

\section{Relaxation in The V-Shape Potential}
\label{App:PdfVp}

Here, we derive the transient relaxation of the PDF in a V-shape potential after a potential quench.
We use a similar Green's function approach to the one used in the case of SR.
We denote the spread of the distribution by $\alpha\equiv b/k_BT$. For the initial force constant, we will denote the spread of the distribution by $\alpha_i$ and for the final one by $\alpha_f$. Therefore, the initial distribution reads $P_{eq}(x)=\frac{\alpha_i}{2}e^{-\alpha_i|x|}$. The Green's function is given by \cite{Chvosta_2003}
\begin{widetext}
\begin{equation}\label{eq: green function V-shpe}
\begin{split}
    g_V(x,t|y)=&\frac{\Theta(xy)}{2\sqrt{\pi Dt}}\exp{\left[-\frac{(|x|-|y|+\alpha_fDt)^2}{4Dt}\right]}+\frac{\Theta(-xy)}{2\sqrt{\pi Dt}}\exp{(-\alpha_f|x|)}\exp{\left[-\frac{(|x|+|y|-\alpha_fDt)^2}{4Dt}\right]}+\\
    &+\frac{\alpha_f}{4}\exp{(-\alpha_f|x|)}\text{erfc}\left(\frac{|x|+|y|-\alpha_fDt}{2\sqrt{Dt}}\right),
\end{split}
\end{equation}
where $\Theta(x)$ is the Heaviside function and $\text{erfc}(x)$ is the complementary error function. By integrating it over the initial condition, one can find the relaxation of the distribution between two V-shaped potentials,
\begin{equation}
\begin{split}
    f_V(x,t)=&\frac{\alpha_i}{2}\int_{-\infty}^\infty e^{-\alpha_i|y|}g_V(x,t|y)\, dy=\\
    =&\left(\frac{\alpha_i}{4}-\frac{\alpha_f}{4}\right)e^{\alpha_i(\alpha_i-\alpha_f)Dt}e^{(\alpha_i-\alpha_f)|x|}\text{erfc}\left[\alpha_i\sqrt{Dt}+\frac{|x|-\alpha_fDt}{2\sqrt{Dt}}\right]+\frac{\alpha_f}{4}e^{-\alpha_f|x|}\text{erfc}\left[\frac{|x|-\alpha_fDt}{2\sqrt{Dt}}\right]+\\
    &+\frac{\alpha_i}{4}e^{\alpha_i(\alpha_i-\alpha_f)Dt}e^{-\alpha_i|x|}\text{erfc}\left[\alpha_i\sqrt{Dt}-\frac{|x|+\alpha_fDt}{2\sqrt{Dt}}\right]
\end{split}
\label{eq:PdfVp}
\end{equation}
\end{widetext}

This result is shown on Fig~\ref{fig:HistCompExp} together with measured PDF for both compression (panel b) and expansion (panel c).

\section{Relaxation of the Variance in the V-Shape Potential}
\label{App:VarianceVp}

By computing the second moment of the distribution in Eq.~(\ref{eq:PdfVp}) and identifying the slowest decaying transient, we obtain at long times for $\alpha_f>2\alpha_i$
\begin{equation}
    \langle X^2(t)\rangle=\frac{2}{\alpha_f^2}+\frac{2 \alpha_f}{\alpha_i\Delta \alpha}\left(\frac{1}{\alpha_i}-\frac{1}{\Delta \alpha}\right)e^{-\Delta \alpha~\alpha_iDt},
\end{equation}
where $\Delta \alpha = \alpha_f - \alpha_i$.
Therefore the relaxation of the variance after a potential quench is slower than the relaxation after a resetting rate increase, which is characterized by an exponential relaxation at a rate of $\alpha_f^2D$.

\section{Finite-time resetting}
\label{App:CritTime}
If each resetting event takes a finite time, then it needs to be compared to the acceleration gained with respect to the potential quench protocol.

\begin{figure}[htb!]
    \centering
    \includegraphics[width=0.9\linewidth]{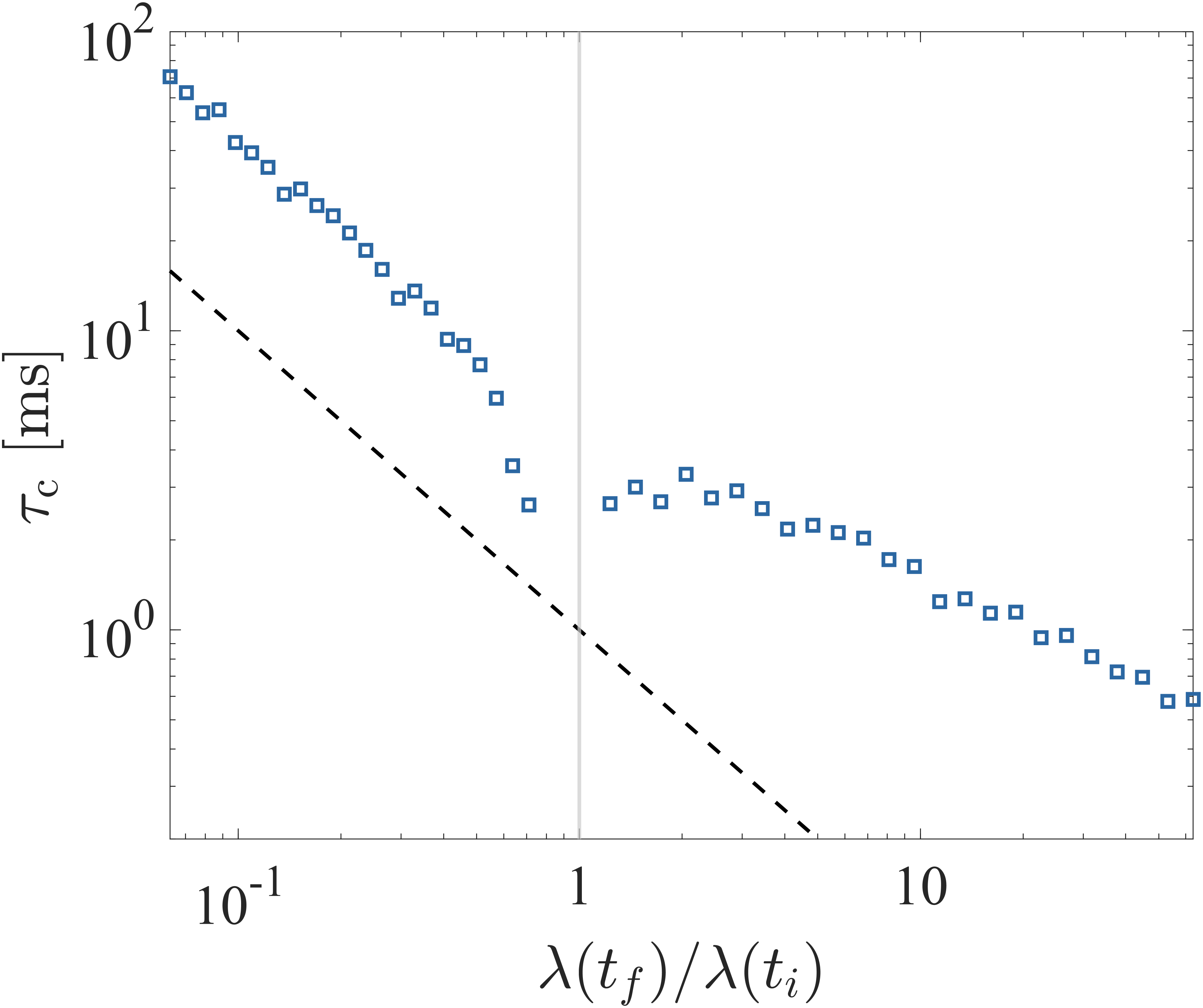}
    \caption{
    Maximal time (in milliseconds) allowed per resetting event to still obtain a net acceleration with respect to a potential quench. It is plotted as a function of the final resetting rate $\lambda_f$ with constant $\lambda_i = 200$ Hz.
    The black dashed line corresponds to $\lambda(t_i) / \lambda(t_f)$
    }
    \label{fig:CriticalTime}
\end{figure}

More precisely, finite-time SR stays beneficial only if the accelerated relaxation time $\tau_{\rm{SR}}$ plus the time needed for each resetting events is still smaller than $\tau_r$.
\begin{equation}
    t_c = \frac{1}{\lambda} \left( \frac{\tau_r}{\tau_{\rm{SR}}} - 1\right)
\end{equation}
For a resetting rate $\lambda$, we can define $N \equiv \lambda * \tau_{\rm{SR}}$ as the average number of resetting events needed during the transient.
Having in mind the protocol presented in the paper, this corresponds to the minimal (average) number of resetting to accelerate a transition between equilibrium states.
If each resetting events takes a time $t$, the finite-time SR stays beneficial as long as $\tau_{\rm{SR}} + N t \leq \tau_r$ which allow to define a critical maximal time per resetting events

On Fig.~\ref{fig:CriticalTime} we plot the critical time $t_c$ as a function of the final resetting rate, with constant initial $\lambda_i = 200$ Hz.
$t_c$ ranges from hundreds of milliseconds for small $\lambda_f$ (i.e. for large decompression with very long relaxation rates) down to a millisecond for large $\lambda_f$ (i.e. strong compression with very short relaxation time).
On all this range, the experimental feasibility of such acceleration is clear.
Indeed, by inducing resetting via intermittent strong optical traps, the relaxation times implied and hence the time needed for each resetting event can be shorter than a millisecond \cite{goerlich2023}.

\newpage
  
%\bibliography{SrAsESE}
%
\end{document}